%% file: BronselaerAcosta2022.tex
\documentclass[12pt]{elsarticle}

\usepackage{algpseudocode}
\usepackage{algorithm}
\usepackage{multirow}
\usepackage{graphicx}
\usepackage{amsmath}
\usepackage{amssymb}
\usepackage{url}
\usepackage{todonotes}
\usepackage{hyperref}
\usepackage{tikz}
\usepackage{pgfplots}
\usepackage{subcaption}
\usepackage{dsfont}
\usepackage{booktabs}
\usepackage{makecell}

\usepackage{stmaryrd}
\usepackage{theorem}
\theoremstyle{break}
\newtheorem{definition}{Definition}
\newtheorem{theorem}{Theorem}
\newtheorem{proposition}{Proposition}

\newproof{proof}{Proof}

\newcommand{\schema}{\mathcal{R}}
\newcommand{\rel}{R}

\newcommand{\attr}{\mathcal{A}}

\newcommand{\fd}{\phi}

\newcommand{\kpart}{\mathcal{K}}
\newcommand{\kvalue}{\mathbf{k}}

\newcommand{\rul}{E}
\newcommand{\rules}{\mathcal{E}}

\newcommand{\invol}[1]{\mathcal{I}\left(#1\right)}
\newcommand{\suf}[1]{\underline{\Omega}\left(#1\right)}
\newcommand{\sol}{\mathcal{S}}
\newcommand{\cover}{\mathcal{C}}

\newcommand{\repair}{\rel^*}
\newcommand{\reptup}{r^*}
\newcommand{\cost}{\Delta}
\newcommand{\icost}{\Delta^\bullet}
\newcommand{\idom}{A^\bullet}
\newcommand{\lcost}[1]{\cost_{#1}}
\newcommand{\indlcost}[1]{\lcost{#1}^\bullet}
\newcommand{\clas}{\rel_{\kpart=\kvalue}}
\newcommand{\lcf}[1]{r_{(#1)}}

\makeatletter
\def\ps@pprintTitle{%
  \let\@oddhead\@empty
  \let\@evenhead\@empty
  \def\@oddfoot{\reset@font\hfil\thepage\hfil}
  \let\@evenfoot\@oddfoot
}
\makeatother

\begin{document}

\begin{frontmatter}


\title{Consistent data fusion with Parker}

\author[antoon]{Antoon Bronselaer\corref{cor1}}
\cortext[cor1]{Corresponding author}
\ead{antoon.bronselaer@ugent.be}
\address[antoon]{DDCM Lab, Department of Telecommunication and Information Science, Ghent University, Sint-Pietersnieuwstraat 41, B-9000 Ghent, Belgium}

\author[maribel]{Maribel Acosta}
\ead{maribel.acosta@rub.de}
\address[maribel]{Faculty of Computer Science, Ruhr University Bochum, Universit\"atsstraße 150, D-44801 Bochum, Germany}

\begin{abstract}
When combining data from multiple sources, inconsistent data complicates the production of a coherent result.
In this paper, we introduce a new type of constraints called edit rules under a partial key (EPKs).
These constraints can model inconsistencies both within and between sources, but in a loosely-coupled matter.
We show that we can adapt the well-known set cover methodology to the setting of EPKs and this yields an efficient algorithm to find minimal cost repairs of sources.
This algorithm is implemented in a repair engine called Parker.
Empirical results show that Parker is several orders of magnitude faster than state-of-the-art repair tools.
At the same time, the quality of the repairs in terms of $F_1$-score ranges from comparable to better compared to these tools.
\end{abstract}

\begin{keyword}
Data Fusion \sep Data Quality \sep Edit rules \sep Key constraints \sep Functional Dependencies
\end{keyword}
\end{frontmatter}


\section{Introduction}
We study the data fusion problem where data from different sources must be under certain consistency constraints.
This problem is modelled by assuming that each data source has the same schema and provides information on some entity in terms of single tuples.
Consistency is validated by means of two types of integrity constraints.
On one hand, tuples are subject to tuple-level constraints that are expressed as \emph{edit rules}.
These edit rules model the internal consistency for each individual source.
On the other hand, tuples coming from different sources must agree on certain attributes and this is modelled as a (partial) key constraint.
This partial key constraint models the level agreement that sources must reach in order to be mutual consistent.

\begin{figure}[ht!]
\centering
\includegraphics[width=1\columnwidth]{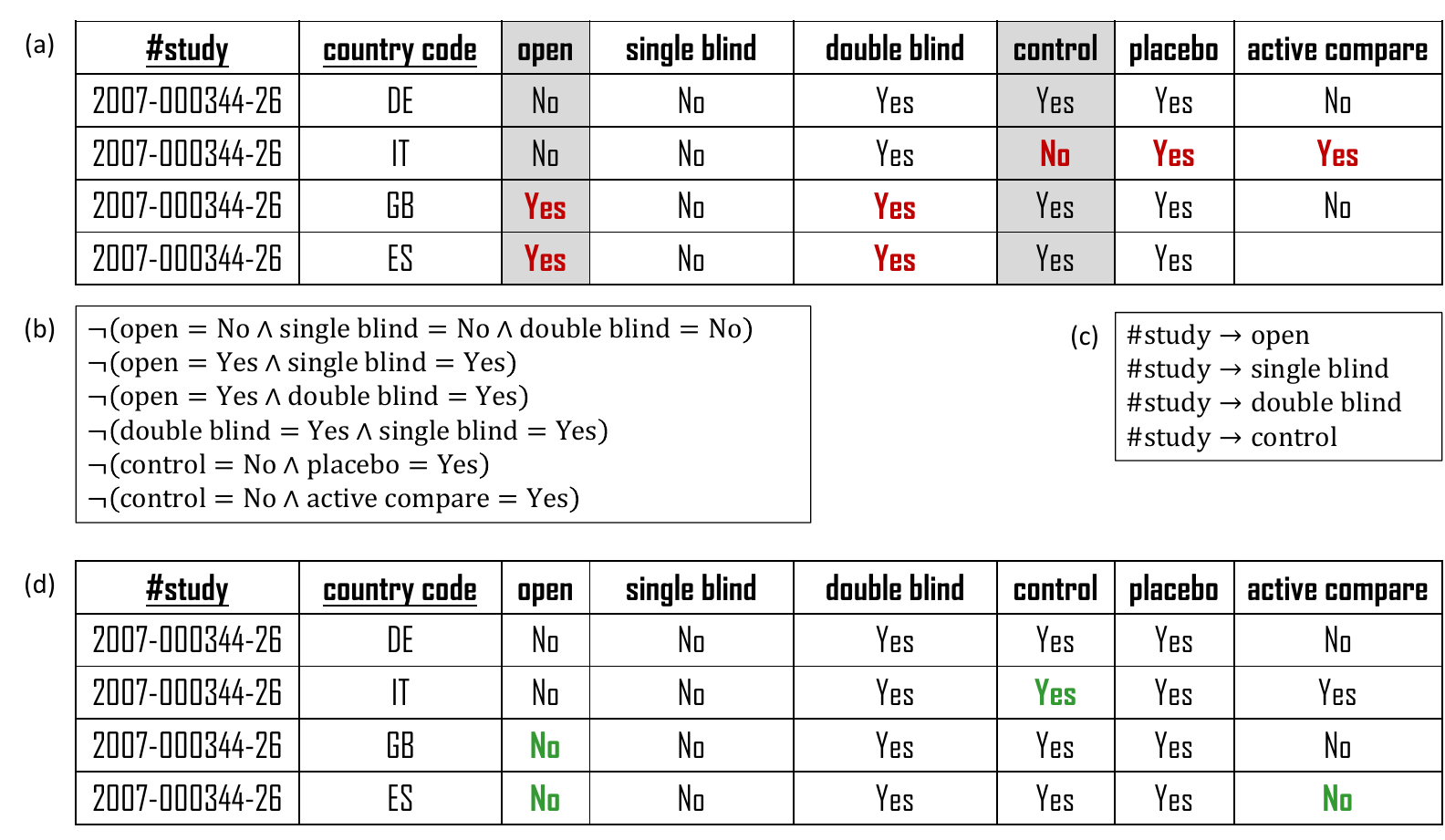}
\caption{Data on the design parameters of a clinical trial executed in different sites (a). Edit rules that need to hold are shown in (b) and violations are marked in bold red font. Required agreement between sites is modelled by a partial key (c), with violations marked in grey. A minimal repair is shown in (d), with changes marked in bold green font.}
\label{fig:running-example}
\end{figure}

\paragraph{Running example} Figure~\ref{fig:running-example} shows data taken from the \href{https://www.clinicaltrialsregister.eu/ctr-search/search}{European Clinical Trials Register} (EudraCT) database.
In Figure~\ref{fig:running-example} (a), the design parameters of a single clinical trial are shown for different sites in the European Union where this trial was executed.
Figure~\ref{fig:running-example} (b) shows six simple edit rules (ERs) to which the design parameters of a study must adhere.
The first four rules together enforce that a study has exactly one masking strategy.
The last two rules enforce that if a trial uses a placebo or an active comparator, then that trial is controlled.
Violations of these rules are shown in bold red font in panel (a).

Figure~\ref{fig:running-example} (c) shows a set of functional dependencies (FDs) that indicate `\#study' is a partial key as `\#study' and `country code' are a candidate key.
The partial key implies that four design parameters need to be the same for the same study, regardless of the site where it was executed.
Violations of this partial key constraints are marked as grey shaded columns in Figure~\ref{fig:running-example} (a).
Finally, Figure~\ref{fig:running-example} (d) shows a \emph{repair} of the data from Figure~\ref{fig:running-example} (a) that violates no ERs and no FDs.
This repair is \emph{minimal} in the sense that there is no relation with the same attributes that (i) satisfies all constraints and (ii) differs from Figure~\ref{fig:running-example} in less attribute values.

\paragraph{Motivation} An immediate question might be why constraints are modelled as a mixture of FDs and ERs.
All constraints in Figure~\ref{fig:running-example} can be represented in the 
9
 framework of conditional functional dependencies (CFDs) \cite{Bohannon2007,Fan2008} or denial constraints (DCs) \cite{Chu2013}, so why not use them?
The main problem with these formalisms, is that finding (minimal) repairs is a computationally intensive task \cite{Geerts2013,Geerts2019,Rekatsinas2017,Mahdava2020}.
When using edit rules, however, the problem of finding (minimal) repairs boils down to finding (minimal) set covers of failing rules if the given rules satisfy a closure property \cite{Fellegi1976, Boskovitz2008, DeWaal2011}.
In a similar manner, finding (minimal) repairs for violations of a partial key constraint is easy if no backward repairing (i.e., modification of the key values) is allowed.
The main argument of this paper is that when combining both partial key constraints and edit rules, we can exploit the properties of both constraint types to efficiently prune the search space and still ensure minimal repairs.
Moreover, the combination of partial keys and edit rules alleviates the expressiveness of edit rules in such a way we can cover many constraints we observe in real life scenarios.
As such, an appealing balance between computational efficiency and expressiveness is obtained.

\paragraph{Contributions} We propose an efficient algorithm to find all minimal repairs when constraints are a mixture of a partial key and a set of edit rules.
The algorithm is implemented as the Parker (Partial Key and Edit Rule) repair engine and is available as a part of an open source framework\footnote{\url{https://gitlab.com/ledc/ledc-sigma}}.
Empirical results show that Parker is superior on all tested datasets in terms of performance when compared to two state-of-the-art cleaning tools (HoloClean and Raha-Baran).
In terms of precision and recall, Parker is competitive with these tools and in some cases better.

The remainder of the paper is structured as follows.
Section~\ref{sec:prelim} explains preliminary concepts. 
Section~\ref{sec:main} introduces the foundations of edit rules under partial key constraints (EPKs). 
Section~\ref{sec:minimal_repairs} describes our approach to compute minimal repairs for EPKs based on the cost models presented in Section~\ref{sec:cost-models}. 
Related Work is discussed in Section~\ref{sec:related-work}, while Section~\ref{sec:experiments} presents an experimental study against the state of the art. 
Lastly, Section~\ref{sec:conclusion} presents the conclusion and an outlook to future work.     

\section{Preliminaries}
\label{sec:prelim}
\subsection{The relational model}
We consider $\attr$ to be a countable set of attributes where for each $a\in \attr$, $A$ denotes the \emph{domain} of $a$ and is assumed to be countable.
A \emph{schema} $\schema=\{a_1,\ldots,a_k\}$ is a non-empty and finite subset of $\attr$.
A \emph{relation} $\rel$ with schema $\schema$ is defined by a finite set $\rel\subseteq A_1\times \ldots \times A_k$.
Elements of $\rel$ with schema $\schema$ are called \emph{tuples} with schema $\schema$.
For a relation $\rel$ with schema $\schema$, a set of attributes $X\subseteq\schema$ and a predicate $P$, we denote the \emph{projection} of $\rel$ over $X$ by $\rel[X]$ and \emph{selection} over $P$ by $\rel_{P}$.
For two relations $\rel'$ and $\rel''$ with respective schemas $\schema'$ and $\schema''$, $\rel'\bowtie\rel''$ denotes the \emph{natural join} of $\rel'$ and $\rel''$.

\subsection{Functional dependencies and partial keys}
A \emph{functional dependency} (FD) \cite{Abiteboul1995} $\fd$ defined over a schema $\schema$ is an expression of the form $X\rightarrow Y$ such that $X\subseteq \schema$ and $Y\subseteq \schema$.
A relation $\rel$ with schema $\schema$ \emph{satisfies} $\fd$ (denoted by $\rel\models \fd$) if $\forall r_1, r_2\in \rel: r_1[X]=r_2[X] \Rightarrow r_1[Y]=r_2[Y]$. 
An FD $X\rightarrow Y$ is \emph{minimal} is there is no $X'\subset X$ such that  $X'\rightarrow Y$ holds.
A set of attributes $\kpart\subseteq\schema$ is called a \emph{candidate key} if the FD $\kpart\rightarrow \schema$ holds and is minimal.
For a candidate key $\kpart$, a partial key is a subset of that candidate key that determines at least one attribute other than the key. 

\subsection{Edit rules}
An \emph{edit rule} \cite{Fellegi1976} is a tuple-level constraint that specifies which combinations of attribute values are not permitted to occur together.
There are different types of edit rules such as \emph{linear} edit rules \cite{Fellegi1976, DeWaal2011}, \emph{ratio} edits \cite{DeWaal2011} and \emph{constant} edits \cite{Boskovitz2008}.
For simplicity, we will restrict ourselves to constant edit rules, even though the methodology is applicable to a broader category of constraints\footnote{The implementation in \href{http://gitlab.com/ledc/ledc-sigma}{ledc-sigma} allows rules equivalent to relational selection $\sigma$ with variables.}.

A (constant) \emph{edit rule} (ER) $\rul$ on $\schema=\{a_1,...,a_k\}$ is an expression of the form $E_1 \times ... \times E_k$ where $E_i\subseteq A_i$.
A tuple $r$ \emph{satisfies} $\rul$ (denoted by $r\models \rul$) if $r\notin \rul$.
If some $\rul_i$ is empty for $\rul$, then $\rul$ never fails and we call such rules \emph{tautologies}.
Conversely, if all $\rul_i$ equal $A_i$ for rule $\rul$, then $\rul$ is never satisfied and we call such rules  \emph{contradictions}.
In what follows, we assume that rules are neither a tautology nor a contradiction.
An attribute $a_i\in\schema$ is said to \emph{enter} an edit rule $E$ if and only if $E_i\subset A_i$.
Alternatively, we say that $E$ \emph{involves} $a_i$.
The set of attributes involved in $\rul$ is denoted by $\invol{\rul}$.

\section{Edit Rules under Partial Key Constraints}
\label{sec:main}

\subsection{Basic definition and problem statement}
Suppose we have a set of data sources that all provide information on some entity.
They do so by each emitting one tuple that obeys a common schema $\schema$.
This means that we can model the whole of these different sources as a single relation $\rel$ with schema $\schema$.
In our earlier example (Figure~\ref{fig:running-example}, (a)), the sources correspond to member states of the EU and they each present information about one clinical trial identified by its study number.
Suppose now we want to fuse information coming from these sources in a consistent manner by enforcing two requirements.
\begin{enumerate}
\item Each source is internally consistent according to a set of tuple-level constraints (Figure~\ref{fig:running-example}, (b)).

\item Sources agree on certain parts of the data and we formalize this by stating that some identifier of the entity we are describing, functionally determines some (not necessarily all) attributes (Figure~\ref{fig:running-example}, (c)).
\end{enumerate}

The combination of these two requirements can be formalized via the notion of an edit rule under a partial key (EPK) constraint and is defined as follows.
\begin{definition}
\label{def:epk}
Let $\schema$ be a schema that can be partitioned into $\kpart$, $\schema_1$ and $\schema_2$.
An edit rule under a partial key constraint (EPK constraint) is an edit rule $\rul$ on $\schema_1 \cup \schema_2$ such that the FD $\fd:\kpart\rightarrow \schema_1$ needs to hold.
\end{definition}
Definition~\ref{def:epk} implies that any EPK is determined by a pair $(\fd,\rul)$.
It thus follows naturally that a relation $\rel$ fails an EPK whenever either $\rel\not\models\fd$ or there is some $r\in\rel$ for which $r\not\models\rul$.
For a single schema $\schema$, multiple edit rules can be defined, but we restrict ourselves to a single partial key.
A set of EPKs can then always be denoted by a couple $\left(\fd, \rules\right)$, where $\fd$ provides the partial key constraint and $\rules$ is a set of edit rules.

The fusion model we introduced above can now be represented by a set of EPKs where
consolidation is done by repairing inconsistencies in an optimal way.
In more formal terms, if $\rel$ fails some EPK, then we want to \emph{modify} $\rel$ into a new relation $\repair$ such that (i) $\repair$ fails no EPKs and (ii) the differences between $\rel$ and $\repair$ are minimized.
A relation $\repair$ is called a \emph{repair} if is satisfies (i) and is called a \emph{minimal repair} if it satisfies (i) and (ii).

In the scope of this paper, we focus on repairs achieved by updates only.
Moreover, we make the explicit assumption that $\kpart$ does not contain errors, modelling the fact that sources make judgements on the right entities.
The minimization criterion of repairs is defined here in terms of a \emph{cost function} for attribute changes.
\begin{definition}
A \emph{cost function} for $a\in\schema$ is a positive-definite function $\lcost{a}:A^2\rightarrow\mathbb{N}$.
\end{definition}
If $\schema$ is entirely equipped with cost functions, then the cost for modifying some tuple $r$ into a tuple $\reptup$ is simple computed by the sum of costs:
\begin{equation}
\cost(r,\reptup) = \sum_{a\in\schema}\lcost{a}\left(r[a], \reptup[a]\right).
\end{equation}
A repair $\repair$ is now called $\cost$-minimal if rows $r\in\rel$ are modified into rows $\reptup$ such that the sum of all $\cost\left(r, \reptup\right)$ is minimal.
One obvious cost function is the one that simply verifies whether $r[a]=\reptup[a]$ (cost $0$), or not (cost $1$).
Using this cost function, $\cost(r,\reptup)$ equals the number of attributes that were modified.
Figure~\ref{fig:running-example} (d) shows a repair of Figure~\ref{fig:running-example} (a) that is $\cost$-minimal under this counting function.

In general, cost functions need not to be symmetric and we can assign a different cost function to each individual tuple.
It is usually a desirable feature that cost functions are at least constant with respect to changing a \textsc{null} value, meaning that $\lcost{a}(\bot,v)$ is independent of $v$.
We now formulate the main problem studied in this paper as follows.
\begin{figure}[ht!]
\centering
\includegraphics[width=0.9\columnwidth]{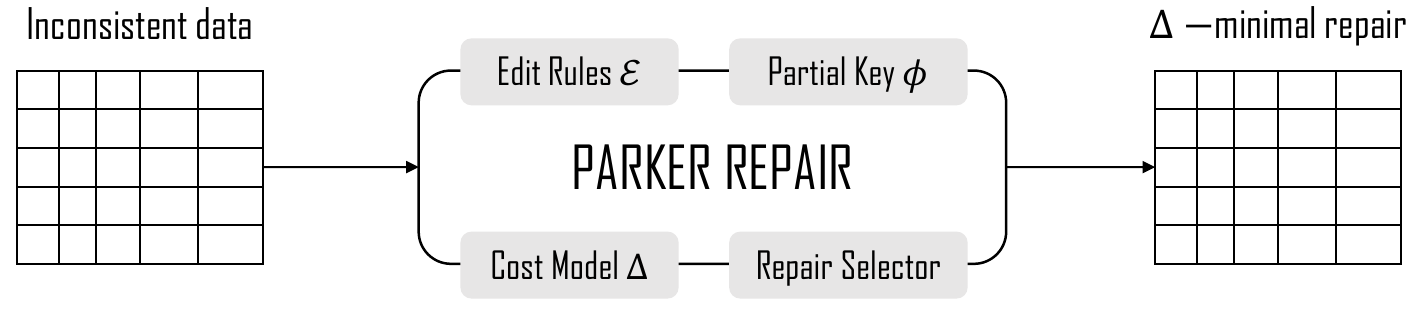}
\caption{High-level architecture of the Parker repair engine.}
\label{fig:repair-engine}
\end{figure}

\paragraph{Problem statement} Given a relation $\rel$ with schema $\schema$ and a set of EPKs, find all $\cost$-minimal repairs $\repair$ for some predefined cost function $\cost$.
Solving this problem in a naive way, requires testing all possible repairs, which has a time complexity of $\mathcal{O}\left(|\rel| \cdot \prod_{a_i\in\schema} \left|A_i\right|\right)$.
By exploiting the properties of EPKs, we are able to construct a repair engine (Parker) that explores the search space in a very efficient manner.
The main architecture of Parker is shown in Figure~\ref{fig:repair-engine}.
Besides the constraints, Parker is equipped with a cost model $\cost$ and a selection strategy to choose, from all $\cost$-minimal repairs, a single repair.
We provide more details on both cost models and selection strategies in Section~\ref{sec:cost-models}.

\subsection{Properties of EPKs}
For constant ERs, an efficient method exists to find minimal repairs \cite{Fellegi1976}.
If some tuple $r\in\rel$ fails some $\rul\in\rules$, a \emph{solution} is a set of attributes $\sol\subseteq\schema$ such that there exists a valid tuple $\reptup$ that \emph{only} differs from $r$ in $\sol$.
In other words, a solution $\sol$ must satisfy:
\begin{equation}
\exists \reptup:\reptup\models\rules \wedge r\left[\overline{\sol}\right]=\reptup\left[\overline{\sol}\right].
\end{equation}
We say that a repair $\reptup$ for $r$ has solution $\sol$ if this above condition is satisfied.
For constant edit rules, solutions satisfy a particular property in the sense that constant edit rules are \emph{interaction-free} in the sense that if we consider a single constant rule $\rul$ and $r$ is changed in two different attributes $a_1$ and $a_2$ to satisfy $\rul$, then one of both changes can be undone and $\rul$ is still satisfied.

To find solutions efficiently, $\rules$ can be transformed into a \emph{sufficient set}, denoted by $\suf{\rules}$.
This sufficient set has the property that any solution $\sol$ for a tuple $r$ is always a set-cover of the \emph{failing} rules from $\suf{\rules}$.
Set covers are hereby defined as follows:
\begin{definition}
A set cover for edit rules $\rules$ on $\schema$ is a set of attributes $\cover\subseteq\schema$ such that $\forall \rul\in\rules: \invol{\rul}\cap \cover \neq \emptyset$.
\end{definition}
In Figure~\ref{fig:running-example}~(c), the set of rules is sufficient, meaning that solutions to errors can be found by composing set covers to failing rules.
For example, the second tuple in Figure~\ref{fig:running-example}~(a) fails two rules and as the attribute `control' covers both of these rules, we are certain there exists repair for the second tuple where we only change attribute `control'.

The construction of $\suf{\rules}$ is key in our approach for several reasons.
First, $\suf{\rules}$ can be constructed efficiently for small to medium-sized $\rules$ by using the FCF algorithm \cite{Boskovitz2008}.
Moreover, continuous research is devoted to making this construction step more efficient.
Second, construction of $\suf{\rules}$ allows to check if there a priori \emph{exists} a tuple that satisfies $\rules$ or not.
If there exists one, $\rules$ is called \emph{satisfiable} and it can be shown that $\rules$ is \emph{satisfiable} if and only if $\suf{\rules}$ does not contain a contradiction.
This translates immediately to EPKs.
\begin{proposition}
\label{prop:existence}
A set of EPKs $\left(\fd,\rules\right)$ is \emph{satisfiable} if $\rules$ is satisfiable.
\end{proposition}
Third, $\suf{\rules}$ allows to find minimal repairs for EPKs, although not in a straightforward manner.
The main problem is that solutions and set covers of failing rules deal with \emph{sets of attributes}.
However, the cost functions $\lcost{a}$ we use here, calculate cost in terms of \emph{values} that attributes take.
We will show later that in the setting of EPKs, we \emph{must} use such cost functions.
The consequence is that we can not simply look for minimal solutions in terms of attributes anymore, because the cost repairing those attributes depends on the values that these attributes take.
We therefore need to rely on a weaker connection between repairs and covers in terms of $\subset$-minimal covers.
\begin{definition}
For edit rules $\rules$ on $\schema$, a set-cover $\cover\subseteq\schema$ is $\subset$-minimal if no real subset of $\cover$ is a set-cover of $\rules$.
\end{definition}
Intuitively, $\subset$-minimal covers inform us about the different ways in how we can change tuples to satisfy failing rules.
In the example on Figure~\ref{fig:running-example} (a), the second tuple fails two rules and we can either solve this by changing `control' or by changing `placebo' and `active compare'.
This intuition leads to the following proposition.
\begin{proposition}
\label{prop:subset-change}
Let $\rules$ be a set of edit rules on $\schema$.
If $\reptup$ is a repair for $r$ with solution $\sol$ then $\sol$ contains at least one $\subset$-minimal cover of the rules failed by $r$.
\end{proposition}
The main question is now whether or not a $\cost$-minimal repair always has a solution $\sol$ that is \emph{equal} to a $\subset$-minimal cover of failing rules of $\suf{\rules}$.
Unfortunately, the answer to this question is negative because, in general, changing more attributes can lead to a lower overall cost.
This means we can not restrict ourselves to $\subset$-minimal covers when searching for $\cost$-minimal repairs.
Fortunately, some properties exist to keep searching efficiently.
For a tuple $r$, a repair $\reptup$ is said to be \emph{$\sol$-minimal} if it has changes only in $\sol$ and does so in a cost-minimal way.
This latter means there is no repair $r'$ with solution $\sol$ such that $\cost(r, r') < \cost(r, \reptup)$.
Two important properties now relate such $\sol$-minimal repairs to $\cost$-minimal repairs and help in finding the latter efficiently.
\begin{theorem}
\label{theorem:boundedness}
If $\reptup$ is a $\cost$-minimal repair for $r$ with solution $\sol$ and
there exists a $\subset$-minimal cover $\cover$ such that $\cover \subset \sol$ then for the tuple $r'$ satisfying $r'[\cover]= \reptup[\cover] \wedge r'[\overline{\cover}] = r[\overline{\cover}]$ we have that
(1) $r'$ is not a repair for $r$ and (2) $\cost\left(r,r'\right)$ is strictly smaller than the cost for any $\cover$-minimal repair for $r$.
\end{theorem}
The crux of Theorem~\ref{theorem:boundedness} is that it gives us a strategy for finding $\cost$-minimal repairs.
More precisely, we can use all $\subset$-minimal cover of failing rules as a starting point.
For any such $\cover$, we can iterate over possible values in order of \emph{lowest cost first}.
We continue this iteration until we find a repair and due to the nature of iteration, this repair is then $\cover$-minimal.
Theorem~\ref{theorem:boundedness} now says that any potential $\cost$-minimal repair $\reptup$ that changes \emph{more} attributes than $\cover$, \emph{must} take values for $\cover$ that have lower cost than the repair we just found.
As such, if we find repairs that are $\cover$-minimal, we only need to inspect those values for $\cover$ that have a strictly lower cost than the $\cover$-minimal solutions and those are values we already iterated over.
A stronger result is that not each combination of values for $\cover$ with cost strictly lower than the $\cover$-minimal solutions needs to be inspected.
\begin{proposition}
\label{prop:necessity}
If $r$ is a tuple that has a $\cost$-minimal repair $\reptup$ using solution $\sol$ then we must have:
\[\forall a_i \in \sol : r[a_i]\neq \reptup[a_i]\wedge \exists \rul\in\suf{\rules}: \reptup[a_i] \notin \rul_i \]
\end{proposition}
Proposition~\ref{prop:necessity} basically states we should only make changes to $r$ that are necessary to satisfy some rule.
Because constant edit rules are interaction-free, Proposition~\ref{prop:necessity} can be verified by observing \emph{only} the values for $\cover$.
This is particularly useful because Theorem~\ref{theorem:boundedness} tells us that if some solution $\sol$ strictly contains a $\subset$-minimal cover $\cover$, then the values taken for $\cover$ by some $\cost$-minimal repair will themselves not resolve all failing rules.
For constant edit rules, we can now also verify whether the values for $\cover$ satisfy Proposition~\ref{prop:necessity}.
We emphasize here that this latter observation holds \emph{only} for constant edit rules.
It one wishes to use edit rules involving variables (e.g. a rule of the type $a_1\leq a_2$), then because such rules are not interaction-free, we can not apply this property if we want to find $\cost$-minimal repairs.

\section{Finding Minimal Repairs for EPKs}
\label{sec:minimal_repairs}
In this section, we exploit the properties of edit rules and EPKs introduced in the previous to compose an efficient algorithm for finding all $\cost$-minimal repairs for a relation $\rel$ constrained under a set of EPKs $(\fd,\rules)$.
To gradually introduce our approach, we will first study the case where $\fd$ is a \emph{full} key constraint.
In that scenario, we have $\fd:\kpart\rightarrow\schema_1\cup\schema_2$ and $\rules$ contains edit rules defined over $\schema_1\cup\schema_2$.
We will introduce our algorithm in that case step by step.
Finally, we will generalize that algorithm to the case of partial keys.

\subsection{Finding $\equiv_\kpart$ tuples}
To begin with, $\kpart$ imposes an equivalence relation $\equiv_\kpart$ on $\rel$ where $r\equiv_\kpart r' \Leftrightarrow r[\kpart]=r'[\kpart]$.
It is easy to see that (i) finding $\cost$-minimal repairs requires treating all equivalent tuples at the same time and (ii) tuples that are not equivalent can be treated independently from each other.
For that reason, we focus here on the treatment of a single equivalence class $\clas$ with $\kvalue$ some key value.
Semantically speaking, each class forms a group of information about a single entity provided by different sources.

\subsection{Minimal cost iteration}
If we have some class $\clas$ then our main interest lies in the attributes $\schema_1\cup\schema_2$ which we can write as a set $\{a_1,\ldots,a_m\}$.
Any $\cost$-minimal repair must take values for $\schema_1\cup\schema_2$ in the set $A_1\times\ldots\times A_m$.
Iterating over all these values this requires testing $\prod_{i=1}^m |A_i|$ combinations which rapidly becomes intractable.
To make this more feasible, we can compose a ranking over the set $A_1\times\ldots\times A_m$ such that lower ranked tuples imply a lower cost when using them as a repair.
In other words, this ranking allows us to iterate over possible repairs in order of \emph{lowest cost first} (LCF).

In order to construct such an LCF ranking, we first consider weak orders $\leq_i$ over the different attribute domains $A_i$.
For each attribute $a\in \schema_1\cup\schema_2$ and for any value $v\in A$, we can see that if $\reptup$ would be a repair for the class $\clas$ with $\reptup[a] = v$, then the cost contributed by $a$ is equal to:
\begin{equation}
c\left(v\right) = \sum_{r'\in \clas} \lcost{a}(r'[a], v)    
\end{equation}
We hereby simply sum the costs required to change the value of $a$, for each member of the class $\clas$ into $v$.
The total cost for repairing class $\clas$ with tuple $\reptup$ is then equal to the sum of all $c(\reptup[a])$ over $\schema_1\cup\schema_2$.
In other words, we have that:
\begin{equation}
\beta(\kvalue, \reptup) = \sum_{r\in\clas} \cost(r,\reptup) = \sum_{a\in\schema_1\cup\schema_2} c\left(\reptup[a]\right)
\end{equation}
We use the notation $\beta\left(\kvalue, .\right)$ here to indicate the local objective we want to minimize for each class $\clas$.
In the example of Figure~\ref{fig:running-example} (a), suppose we consider a constant cost model $\lcost{a}$ for each attribute $a$ where $\lcost{a}(v,v')$ is $0$ if $v=v'$ and $1$ if $v\neq v'$.
In that case, for attribute `open' we find $c(\text{`No'})=c(\text{`Yes'})=2$.
Similarly, for attribute `control', we find $c(\text{`No'})=1$ and $c(\text{`Yes'})=3$.
Note here that, although the cost model $\cost_a$ is constant for individual rows, the \emph{induced} cost for changing some value for \emph{an entire class}, is not.

Consider now a weak order $\leq_a$ over $A$ such that $v\leq_a v'$ if and only if $c\left(v\right)\leq c\left(v'\right)$.
Next, we choose a \emph{total} order that is \emph{consistent} with $\leq_a$ by 
\emph{indexing} values from $A$ in such a way that lower indices imply lower cost.
In other words, we consider an index scheme such that $\forall v_{(i)}\in A:\forall v_{(j)}\in A: i\leq j \Leftrightarrow v_{(i)}\leq_a v_{(j)}$, where both $i$ and $j$ are values in $\{1,\ldots,|A|\}$.
With these notations at hand, Algorithm~\ref{algo:tuple-iteration} provides an algorithm for LCF-iteration over values from $A_1\times\ldots\times A_m$.

\begin{algorithm}
\caption{LCF-tuple iteration}
\label{algo:tuple-iteration}
\footnotesize
\begin{algorithmic}[1]
\Require $\left\{(A_1, \leq_1), \ldots(A_m, \leq_m)\right\}$
\State $\mathbb{S} \gets\emptyset$ \label{line:init-empty-set}
\State \textbf{push}$\left(\mathbb{S}, \left[ v^1_{(1)},\ldots,v^k_{(1)}\right]\right)$ \label{line:first-push}
\While{$\mathbb{S} \neq \emptyset$}
\State $r_\text{pop}\gets$\textbf{pop}$\left(\mathbb{S}\right)$ \label{line:pop}
\For{$i\in \{1,\ldots,k\} $}
\If{$r_\text{pop}[a_i] \neq v^i_{(|A_i|)}$} \label{line:index-check}
\State \textbf{push}$\left(\mathbb{S},\textbf{next}\left(r_\text{pop}, i\right)\right)$ \label{line:re-push}
\EndIf
\EndFor
\EndWhile
\end{algorithmic}
\end{algorithm}

The algorithm takes a set of $m$ weakly-ordered sets $(A_i,\leq_i)$, for which we can generate indexed values consistent with $\leq_i$.
The main idea of the iteration is to keep a \emph{sorted set} $\mathbb{S}$ that keeps tuples $r'$ sorted by their value $\beta(\kvalue,r')$.
Initially, we push the value with indices one and we sequentially pop the first element from $\mathbb{S}$.
For each popped tuple $r_\text{pop}$ we check, for each attribute $a_i$, if $r_\text{pop}$ currently has the last value according to index scheme (line~\ref{line:index-check}).
If it does not, we push a copy of $r_\text{pop}$, but increase the index of the value for $a_i$ with one step (line~\ref{line:re-push}).

Algorithm~\ref{algo:tuple-iteration} can be attributed the following properties.
First, sorted sets like $\mathbb{S}$ can be implemented with a Red-Black tree, meaning that inserting new tuples $r'$ is done in $\mathcal{O}(\log |\mathbb{S}|)$ time.
Second, after each pop, $\mathbb{S}$ grows with \emph{at most} $m$ tuples, where $m$ equals the amount of attributes.
Third, each tuple we push to $\mathbb{S}$ has a value for $\beta$ that is greater than or equal to the value for $\beta$ of the tuple that was popped.
Because we maintain order in $\mathbb{S}$, it follows that we pop values from $\mathbb{S}$ in such a way that $\beta$ is never decreasing.
Lastly, by construction, each value from $A_1\times\ldots\times A_m$ must be pushed once to $\mathbb{S}$, which means the sequence of popped values is indeed an iteration over $A_1\times\ldots\times A_m$.
Note that it is possible that the same tuple is popped multiple times from $\mathbb{S}$ but this can be solved by either popping all tuples with an equal value for $\beta(.,.)$ in one go or by keeping track of those tuples we already observed.

The properties we mentioned imply that finding minimal repairs of class $\clas$ can be done by simply running Algorithm~\ref{algo:tuple-iteration} and check, for each popped $r'$ whether $r'\models \rules$.
If so, we found one minimal solution and it suffices to verify only solutions with an equal cost.
As soon as we observe solutions with higher cost, we can stop the algorithm and return all tuples that satisfy $\rules$.
Clearly, this search still has a worst case complexity of $\mathcal{O}\left(\prod_{i=1}^m |A_i|\right)$, but if \emph{consistent} tuples (i.e., tuples that satisfy $\rules$) are observed early, we do much better than this worst case.
Still, in this naive strategy, we will in general observe tuples for which we know they will not be consistent and skipping those tuples can further improve the efficiency of the search.

\subsection{Exploiting independence of rules}
One way of improving the complexity of LCF iteration is to treat groups of attributes independently.
Suppose we can partition $\rules$ in two sets $\rules_1$ and $\rules_2$ such that attributes involved in rules from $\rules_1$ are not involved in rules from $\rules_2$ and vice versa.
Formally:
\begin{equation}
\left(\bigcup_{\rul\in\rules_1}\invol{\rul}\right) \cap \left(\bigcup_{\rul\in\rules_2}\invol{\rul}\right)= \emptyset. 
\end{equation}
When this condition is satisfied, we say that $\rules_1$ and $\rules_2$ are \emph{independent}.
For a partition into independent sets, there are two essential properties that hold.
First, Barcaroli and Venturi \cite{Barcaroli1997} showed that
if $\rules$ partitions into independent sets $\rules_1$ and $\rules_2$, then 
$\suf{\rules} = \suf{\rules_1} \cup \suf{\rules_2}$.
Second, the following proposition holds.
\begin{proposition}
\label{prop:indep2}
Let $\left(\fd, \rules\right)$ be a set of EPKs on $\schema$ and $\rel$ a relation with schema $\schema$.
If $\rules$ partitions into two independent sets $\rules'$ and $\rules''$ such that $\rules'$ are edit rules on $\schema'\subseteq \schema$ and $\rules''$ are edit rules on $\schema''\subseteq \schema$ with $\schema'\cup\schema'' = \schema\setminus\kpart$ then $\repair$ is a $\cost$-minimal repair if and only if $\repair[\kpart \cup\schema']$ is a $\cost$-minimal repair for $\rel[\kpart \cup\schema']$ against $\left(\fd, \rules'\right)$ and $\repair[\kpart \cup\schema'']$ is a $\cost$-minimal repair for $\rel[\kpart \cup\schema'']$ against $\left(\fd, \rules''\right)$.
\end{proposition}
These results show for any $(\fd, \rules)$ that if $\rules$ is composed of independent sets, then repairing inconsistencies can be done for $(\fd, \rules_1)$ and $(\fd, \rules_2)$ separately.
Note that recursive application of both propositions allows to generalize the results to a partition into $K$ independent sets.
In Figure~\ref{fig:running-example} (b), we have two independent sets: the first four rules and last two rules.

\subsection{Finding minimal repairs under full key constraints}
We now combine all the results from the previous into an algorithm.
The main idea is that, when we perform an LCF iteration over some set $A_1\times\ldots\times A_m$, we try to keep $m$ small by focusing on $\subset$-minimal covers.

Consider some class $\clas$ and let us now index tuples consistent with the LCF order, just as we did for individual attributes.
That is, the first tuple under LCF order is $\lcf{1}$, the second tuple $\lcf{2}$ and so on.
There are now two cases.

$\boxed{Case\ 1}$. Suppose $\lcf{1}\models \rules$, then this tuple is one of the minimal repairs for $\clas$. 
That means, if we replace the attributes $\schema_1\cup\schema_2$ of each tuple in $\clas$ with $\lcf{1}$, then the class is repaired with minimal cost.
Of course, $\lcf{1}$ is not guaranteed to be the only minimal cost repair.
Therefore, consider the set of tuples:
\begin{equation}
M_{\kvalue} = \{r \mid r \in A_1 \times\ldots\times A_k \wedge \beta(\kvalue,r) = \beta(\kvalue,\lcf{1}) \}
\end{equation}
This set contains the first $|M_{\kvalue}|$ tuples under LCF order that all have equal cost.
Clearly, if $\exists r'\in M_{\kvalue}:r'\models\rules$, then the set of minimal cost repairs for $\clas$ is a subset of $M_{\kvalue}$ and it suffices to inspect $M_{\kvalue}$.

$\boxed{Case\ 2}$. If the first case does not apply, we have $\forall r'\in M_{\kvalue}:r'\not\models \rules$.
We then want to change $r'$ into some $r''\notin M_{\kvalue}$.
If we do that by modifying its value for $a$, then:
\begin{equation}
\beta\left(\kvalue, r''\right) = \beta\left(\kvalue, r'\right) + c\left(r''[a]\right) - c\left(r'[a]\right)
\end{equation}
for which we know that $c\left(r''[a]\right) - c\left(r'[a]\right) > 0$ because $r''\notin M_{\kvalue}$.
This can be generalized to:
\begin{equation}
\beta\left(\kvalue, r''\right) = \beta\left(\kvalue, r'\right) + \sum_{a\in \schema_1\cup\schema_2} c\left(r''[a]\right) - c\left(r'[a]\right).
\end{equation}
where we require that $c\left(r''[a]\right)>c\left(r'[a]\right)$ to ensure $r''\notin M_{\kvalue}$.
The above expression shows that the cost for repairing $\clas$ by $r''$ can be written as the cost for $r'$ and the functions $c\left(.\right)$.
But since $r'\in M_{\kvalue}$, we know that $\beta\left(\kvalue, r'\right)$ is already minimal.
Hence, in order to minimize $\beta\left(\kvalue, r''\right)$ we must minimize the second term of the right hand side, conditioned on $r''\models \rules$.
To do so, we consider for each attribute $a\in\schema$ an \emph{induced} cost function  $\indlcost{a}$ on the set $\idom = \{v\mid v\in A\wedge c(v)>r'[a]\}$ that satisfies $\forall v\in \idom: \indlcost{a}\left(r'[a], v\right) = c\left(v\right) - c\left(r'[a]\right)$.
This induced cost function gives the cost to change attribute $a$ of tuple $r'$ accounting for all tuples in $\clas$.
By construction, $\indlcost{a}$ is positive definite on $\idom$ and we can now write:
\begin{equation}
\beta\left(\kvalue, r''\right) = \beta\left(\kvalue, r'\right) + \icost(r, r'').
\end{equation}
It follows that if we find a $\icost$-minimal repair $\reptup$ for $r'$, then clearly, we minimize $\beta\left(\kvalue, r''\right)$ under the condition $r''\models\rules$.
This now leads us the main theorem of this paper.
\begin{theorem}
\label{theor:main}
For any class $\clas\subseteq \rel$ with schema $\schema$, some cost model $\cost$ and a set of EPKs $\left(\fd,\rules\right)$, any $\cost$-minimal repair $\reptup$ is the $\icost$-minimal repair of some tuple $r'\in M_\kvalue$.
\end{theorem}
The importance of Theorem~\ref{theor:main} is that it reduces finding minimal repairs for EPKs $\left(\fd,\rules\right)$ to finding minimal repairs for $\rules$, but with a modified cost model $\icost$.
In the previous section, we already pointed out that this can be done by inspection of $\subset$-minimal covers.
Now, as a final step, we connect this inspection to the notion of LCF iteration.
More precisely, suppose we have some $r'\in M_\kvalue$ for which we have the induced cost model $\icost$.
If $r'\not\models\rules$, we compose the failing rules as $\{\rul\mid\rul\in\suf{\rules} \wedge r' \not\models\rul\}$.
Next, we find all $\subset$-minimal covers $\cover$ of this set of failing rules.
For each such cover $\cover$, we can find a minimal $\cover$-repair by using Algorithm~\ref{algo:tuple-iteration} where $m=|\cover|$.
In other words, we do an LCF-tuple iteration, but consider only attributes in $\cover$.
In general, we can expect $|\cover|$ to be significantly smaller than $|\schema|$.
In a final step, we need to account for the fact that the solution attributes of a $\cost$-minimal repair are not necessarily equal to a $\subset$-minimal cover, but can also be a superset of such a cover.
To account for this, we can use Theorem~\ref{theorem:boundedness} and inspect only those value combinations for $\cover$ that occur in LCF order before any $\cover$-minimal repair.
In addition, if we consider constant edit rules, we can use Proposition~\ref{prop:necessity} to discard any value combinations for $\cover$ where some value changes to attributes in $\cover$ are unnecessary.
All the ideas and properties we have presented can now be summarized in the pseudo code of Algorithm~\ref{algo:full-key-repair}.

\begin{algorithm}
\caption{Full Key Repair}
\footnotesize
\label{algo:full-key-repair}
\begin{algorithmic}[1]
\Require Relation $\rel$ with schema $\kpart\cup\schema'$ and EPKs $\left(\kpart\rightarrow \schema',\rules\right)$
\Ensure Relation $\repair$ that is a $\cost$-minimal repair of $\rel$
\State $\suf{\rules}\gets$\textbf{ FCF}$(\rules)$ \label{line:fcf}
\For{$\kvalue\in \rel[\kpart]$}
\State $\varphi\gets\emptyset$
\State $M_{\kvalue}\gets$ \textbf{LCF}($\kvalue, \cost$) \label{line:min-changes}
\If {$\exists r'\in M_{\kvalue}: r'\models\rules$}
\State $\varphi \gets \{r' \mid r'\in M_{\kvalue}: r'\models\rules\}$ \label{line:fast-stop}
\Else
\For{$r'\in M_{\kvalue}$} \label{line:min-change-loop}
\For{$\cover \in $\textbf{ covers}($r', \suf{\rules}$)} \label{line:covers}
\State $\varphi_\cover\gets$ \textbf{findRepairs}($r', \suf{\rules}, \icost, \cover$) \label{line:cover-repair}
\State $\varphi\gets$ \textbf{merge}($\varphi, \varphi_\cover$) \label{line:merge}
\EndFor
\EndFor
\EndIf
\State $\reptup \gets$ \textbf{select}($\varphi$) \label{line:select}
\State \textbf{apply}($\reptup$, $\clas$) \label{line:apply}
\EndFor
\end{algorithmic}
\end{algorithm}

In Algorithm~\ref{algo:full-key-repair}, we assume that $\rules$ cannot be partitioned into independent sets.
If it can, then Algorithm~\ref{algo:full-key-repair} should applied for each independent set separately.
The algorithm starts with composing a sufficient set $\suf{\rules}$ for the rules $\rules$ (line~\ref{line:fcf}).
It is assumed here that $\suf{\rules}$ does not contain a contradiction.
If is does, then the EPKs are not satisfiable (Proposition~\ref{prop:existence}) and the algorithm should return an error.
Next, we iterate over each key value $\kvalue$ and compose the set $M_\kvalue$ of minimal changes according to $\cost$ (line~\ref{line:min-changes}).
If this set contains tuples that satisfy $\rules$, these tuples have minimal cost and we assign all such tuples in $M_\kvalue$ to the set $\varphi$ (line~\ref{line:fast-stop}).
Else, we iterate over each $r'\in M_\kvalue$ and compose the $\subset$-minimal covers $\cover$ for $r'$ in terms of $\suf{\rules}$ (line~\ref{line:covers}).
For each cover $\cover$, we search for possible minimal repairs of $r'$ using $\cover$ as the attributes we change and where we reason in terms of the induced cost model $\icost$ (line~\ref{line:cover-repair}).
To do so, we iterate over possible values for $\cover$ in LCF order, where cost is now determined by $\icost$.
This LCF iteration can be done by using a modification of Algorithm~\ref{algo:tuple-iteration}, where we stop as soon as we have observed all minimal repairs.
In the same step, we also inspect possible supersets of $\cover$, where we can apply Theorem~\ref{theorem:boundedness}.
That means, when we iterate over values for $\cover$ in LCF order, each value combination that does not lead to a $\cover$-minimal repair is fixed and we search for $\icost$-minimal repairs, but now under the condition that the values for $\cover$ are fixed.
If we deal with constant edit rules, we can also use Proposition~\ref{prop:necessity} to restrict this search.
The possible minimal repairs using $\cover$ (or a superset) are stored in the set $\varphi_\cover$ and this set is merged with the repairs $\varphi$.
That means, if the newly found repairs have an equal cost, then we take the union of both sets.
If $\varphi_\cover$ has a lower cost, then $\varphi$ is replaced with $\varphi_\cover$.
Else, we just keep $\varphi$.
Because of the latter, we can keep at all times an upper bound for the cost of repairing.
As we always iterate in terms of LCF order, we can always abort searching whenever we exceed this upper bound.
This upper bound \emph{must} monotonically decrease during the loop over covers (line~\ref{line:covers}) \emph{and} during the loop over tuples $r'\in M_\kvalue$  (line~\ref{line:min-change-loop}).
When we have inspected all $\subset$-minimal covers $\cover$ and their supersets, variable $\varphi$ will contain all $\cost$-minimal repairs for $\clas$ (Theorem~\ref{theor:main}).
We then select one of the minimal repairs (line~\ref{line:select}) and apply it to each of the tuples in $\clas$ (line~\ref{line:apply}).

\subsection{The case of partial key constraints}
Consider now the general case of a partial key constraint $\fd:\kpart\rightarrow\schema_1$ and where $\rules$ contains edit rules defined over $\schema_1\cup\schema_2$.
The main problem is now that a $\cost$-minimal repair for $\rel[\kpart\cup\schema_1]$ and constraints $(\fd,\rules_1)$ will not necessarily lead us to a $\cost$-minimal repair for $\rel$.
In Figure~\ref{fig:running-example} (c), attributes `placebo' and `active compare' are not determined by `\#study', which means these design parameters depend on the country where the study was executed.

\begin{figure}[ht!]
\centering
\includegraphics[width=0.6\columnwidth]{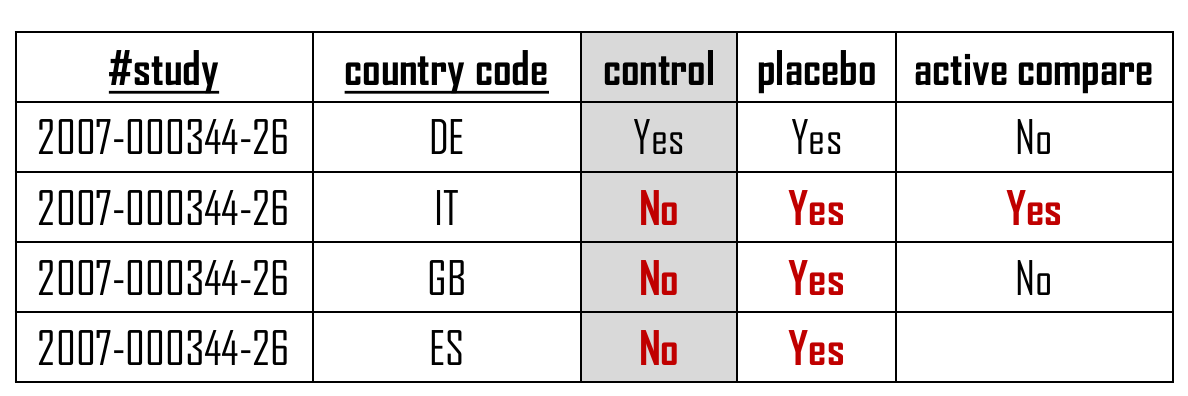}
\caption{A modification of the data in Figure~\ref{fig:running-example}}
\label{fig:example-2}
\end{figure}

In order to solve the more generic problem of partial keys, we can again assume that each set of tuples $\clas$ is treated separately and that there are no independent sets in $\rules$ (if there are, we can use a divide-and-conquer strategy).
In the example of Figure~\ref{fig:running-example}, we have already noted that the set of rules shown in (b) contains two independent sets.
In the first independent set, rules involve only attributes determined by `\#study', so that problem can be treated entirely by Algorithm~\ref{algo:full-key-repair}.
In the second independent set, rules involve both attributes determined by `\#study' (i.e., `control') and attributes not determined by `\#study' (i.e., `placebo' and `active compare').
Let us now consider a slightly modified example as shown in Figure~\ref{fig:example-2}.

In order to find minimal repairs, we first use Algorithm~\ref{algo:full-key-repair} to find a $\cost$-minimal repair for $\clas[\kpart\cup\schema_1]$ under the same cost model but for EPKs $\left(\fd, \rules_1\right)$.
Hereby, $\rules_1$ contains all rules that involve \emph{only} attributes from $\schema_1$.
In the example of Figure~\ref{fig:example-2}, this means we need to find a minimal repair for attribute `control' where $\rules_1=\emptyset$.
Assuming a constant cost model, this leads to a single repair where `control' is assigned the value `No' and cost equals $1$.
If we then consider attributes `placebo' and `active compare', all values `Yes' needs to be changed into `No' to satisfy the edit rules, implying a total cost of $6$.
However, if we would choose `control' equal to `Yes', the total cost would be $3$.

In order to account for this, we can modify our algorithm in a simple manner.
When we search for repairs with induced cost models (Algorithm~\ref{algo:full-key-repair}, line~\ref{line:cover-repair}), we do this by LCF iteration on a space determined by a cover $\cover$.
However, if we find a repair for $\clas[\kpart\cup\schema_1]$, we must then verify if that repair would cause any violations of edit rules for $\clas$.
If it does, we can compute the \emph{additional cost} for fixing those violations for each tuple in $\clas$.
We then keep on progressing the LCF iteration until we are sure we have found all minimal repairs, just like before.
Note that this approach will work best if $|\schema_2|<<|\schema_1|$, which means that most attributes are determined by $\kpart$.
This is a reasonable assumption in the setting of data fusion.

\section{Cost Models and Repair Selection}
\label{sec:cost-models}
After introducing the main methodology for finding repairs, it should be clear that cost functions are of crucial importance in our methodology.
Moreover, as minimal repairs are not unique, we require repair selection methods to select a single repair from the minimal alternatives.
In this section, we discuss some practical methods to build cost functions and repair selection methods.

\subsection{Cost functions}
\label{sec:cost-function-examples}
In the most simple case, a cost function is \emph{constant} and we have:
\[\lcost{a}(v,v')=\left\{\begin{array}{cc}
0  &  v=v'\\
\alpha  &  v\neq v'
\end{array}\right.\]
Here, $\alpha$ is some constant integer cost that can be used to reflect the reliability of attribute $a$.
That is, a more reliable attribute $a$ can be assigned a higher cost $\alpha$.
Using constant cost functions for edit rules has the advantage that all minimal repairs must have a solution equal to a $\subset$-minimal cover, which makes it much more easy to find repairs.
When using constant cost functions in the setting of EPKs, then $\alpha=1$ leads to an induced cost function $\indlcost{a}$ that uses \emph{majority voting} for values.

It has been established in the past that majority voting is not always the best choice as a fusion model \cite{Dong2009,Dong2012}.
A better approach is to account for reliability of sources \cite{Dong2012}, which in our case correspond to tuples in $\rel$.
Usually, reliability is estimated based on some training data.
We propose to deal with reliability in a different, non-supervised manner by assigning each row in $\rel$ it's own cost function.
The constant $\alpha$ we use in a constant cost model is then multiplied with a factor that is inverse proportionate to the `badness' of the row.
This multiplier is computed, for each row $r\in\rel$ as follows:
\begin{equation}
\left(|\schema| - \left|\{a\mid r[a] = \bot \vee \left( a\in\invol{\rul}\wedge r \not\models\rul \right)\}\right|\right)^\omega
\end{equation}
Hereby, we make a very course and pessimistic estimate of the number of attributes that are `in error' by looking at those attributes that either occur in a failing rule or have a null value.
We then subtract that number from the total number of attributes (i.e. $|\schema|$) and raise it to the power $\omega$, which when greater then $1$ acts an amplifier of the effect.
We can then compute the cost for changing the value of attribute $a$ for row $r$ from $v$ to $v'$ as:
\begin{equation}
\lcost{a}(v,v') \cdot \left(|\schema| - \left|\{a\mid r[a] = \bot \vee \left( a\in\invol{\rul}\wedge r \not\models\rul \right)\}\right|\right)^\omega
\end{equation}
By doing so, we basically model that making a change to a row that contains many attributes that might be in error has a relatively low cost.
We will show in Section~\ref{sec:experiments} that this heuristic can provide a significant boost in the quality of repairs.

Besides these two basic cost functions, a plenitude of other cost functions can be considered.
For example, when data are additive, the cost for changing $v$ into $v'$ can account for the \emph{magnitude} $|v-v'|$ of the change.
Another way of dealing with cost, is to account for specific error mechanisms.
For example, when $v$ and $v'$ are string data, one could assign a lower cost to a change where the Levenshtein distance between $v$ and $v'$ is very small, indicating that the change from $v$ to $v'$ can be explained by a typographical error.
In the same spirit, edits of dates could be equipped with error-aware cost functions.
As a final example, a cost function could also account for a priori preferences among values.
Consider for example a case with allergen information.
Assume there are different sources that provide conflicting information on the allergens of a single product.
One source could say that product A contains gluten and another source could say product A does not contain gluten.
In such a scenario, one is inclined to prefer the value of the source that induces the least \emph{risk} and therefore model that the cost of changing `gluten' into `no gluten' is much greater than the cost of the opposite change.
We will demonstrate the impact of such a cost model in Section~\ref{sec:experiments}.

\subsection{Repair selection}
When multiple repairs are found to have a minimal cost, it can be desirable to provide a further selection strategy to choose a `best' repair from the minimal ones.
One obvious method to do so, is simply to pick a minimal repair at random.
In this line of thought, we consider each minimal repair to be equally good and make no further distinction between minimal repairs.

A second method would be to aim at respecting the (marginal and joint) distributions of the observed data as much as possible.
To that end, one often uses the clean data (i.e., the data in $\rel$ that contains no violations of rules) to estimate the distributions of the data.
When such distributions are available, one can select a repair among all minimal repairs with a probability that is proportionate to the observed frequency of that repair.
There are several ways to implement such a strategy.
We will resort in our experiments to a simple method where make an estimate of the probability of observing a repaired object, meaning we measure the frequency of the entire object after repairing.
Alternatives could be to observe the frequency of only the part of the attributes that are changed or make a selection of the attributes for which frequencies are measured.

As a final note, we emphasize the distinction between cost functions and repair selection.
One could argue that repair selection becomes largely disposable if we would include for example information about frequency in the computation of cost.
We do however reject this idea.
In our methodology, \emph{cost} is something that can be used to make \emph{local} distinctions between repairs for a given entity.
It accounts for information about that entity and not other entities.
Whenever there are multiple repairs that are minimal in terms of cost, \emph{global} information about the data, such as distributions, can be used to make a further distinction.
In general, we think this difference is important to avoid an all too large influence of the distributions in for example extremely skewed datasets.

\section{Related Work}
\label{sec:related-work}
Data fusion covers a broad category of problems where information from different sources needs to be consolidated.
In this paper, we study the specific scenario where (i) each source is modelled by a single tuple of a relation and (ii) consolidation means merging into a single tuple for those attributes under the partial key constraint.
This seemingly simple problem, and slight variants of it, has been studied surprisingly often in the vast body of literature dealing with data fusion.
The very first approaches kept a close connection to the relational setting.
They used relational operators like union and (match) join \cite{Legaria1994,Legaria1997, Yan1999, Greco2001}.
Continuing on this idea, the SQL language has been extended with a \textsc{fuse by} operator, yielding a declarative approach towards data fusion \cite{Bleiholder05,Bleiholder10}.
An interesting observation is that these relational approaches usually use operators like subsumption and complement to `clean' a relation and minimize the amount of \textsc{null} values \cite{Bleiholder11}.
In that extent, it can be said that the even earliest approaches used some notion of consistency to optimize the result.

One particular problem with fusion by means of relational operators, is that it does not deal well with errors in data.
To that extent, attribute-level fusion functions can be used to deal with linear conversions \cite{Lu1997,Fan2001}, specificity of information \cite{Bronselaer2016} and multi-valued data \cite{Bronselaer2012,Bronselaer2015}.
A good overview of possible attribute-level fusion functions can be found in \cite{Bleiholder08}.
Another approach to resolve inconsistencies is to add sufficiently many sources.
A particular problem hereby is that sources can copy from one another and knowing these dependencies turns out to be important in source selection \cite{Dong2009}.
Moreover, it has been established that estimating the reliability of the different sources significantly influences the quality of the fusion operator \cite{Dong2012, Li2013}.
In the current paper, we have adopted this idea, but rather than using a training set to estimate reliability, we use an unsupervised estimate of the reliability based on failing rules and missing values.

The attribute-level fusion functions that are mentioned here aim to resolve inter-source inconsistencies.
The approach we develop here is more general in the sense it allows to deal with both intra-source and inter-source inconsistencies.
To that extent, some approaches from the field of constraint-based consistency verification follow the same strategy.
For example, Llunatic allows to model extended equality-generating dependencies (EGDs) and uses a generalization of the Chase algorithm to construct a Chase tree that can be used to search repairs of dirty data \cite{Geerts2013,Geerts2019}.
In practical applications, the branching factor of a Chase tree is so high, that pruning strategies are required to keep the computational effort feasible.
A consequence thereof is that solutions are not guaranteed to be minimal, although there are good heuristics to approximate minimal solutions.
Another way to find repairs, is by learning a probabilistic model of the data \cite{Rekatsinas2017}.
Such a methodology is used in HoloClean, that allows to model a set of denial constraints (DCs) that need to be satisfied on the data \cite{Rekatsinas2017}.
HoloClean uses the available constraints in combination with other information like correlation analysis, outlier detection techniques, reference data... to learn a model for the data.
From that model, it then generates repairs for the given data.
Finally, Raha \cite{Mahdava2019} and Baran \cite{Mahdava2020} are two semi-supervised systems for ``configuration-free'' error detection (Raha) and repair (Baran).
The main idea here is that no constraints need to be given.
Instead, error patterns are learned from a small dirty dataset for which either users provide correct labels or a clean version is available.
Parker differs from these approaches by using a specific type of constraints (EPKs) that are less expressive than extended EGDs and DCs, but can use the set-cover approach to search for minimal repairs efficiently.
Hereby, a wide variety of cost functions can be used to model a broad range of error mechanisms.

\section{Experimental Study}
\label{sec:experiments}

\subsection{Experimental Setup}

In the following, we describe the configurations of our experiments. 
To allow for reproducibility, the datasets, source code, raw results and analyses of the results are available online \footnote{\url{https://gitlab.com/antoonbronselaer/parker-reproducibility}}.

\input{tables/datasets}

\begin{figure}[h!]
\vspace{20mm}
\begin{subfigure}[t]{0.49\textwidth}
\raisebox{8pt}{
\includegraphics[width=\textwidth]{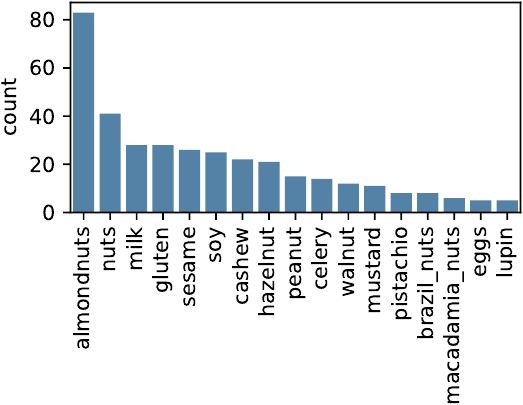}
}
\subcaption[c]{Allergen}
\label{fig:errors_allergen}
\end{subfigure}
\begin{subfigure}[t]{0.29\textwidth}
\raisebox{18pt}{
\includegraphics[width=\textwidth]{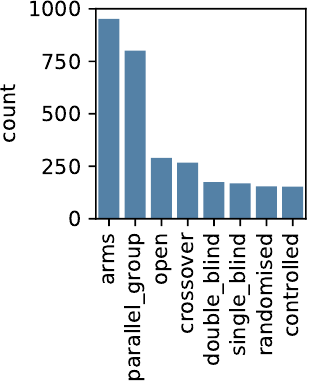}
}
\subcaption[c]{EudraCT}
\label{fig:errors_eudract}
\end{subfigure}
\begin{subfigure}[t]{0.19\textwidth}
\includegraphics[width=\textwidth]{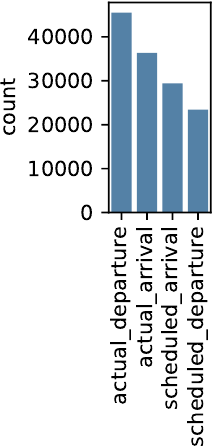}
\subcaption[c]{Flight}
\label{fig:errors_flight}
\end{subfigure}
\caption{Frequency of errors in the different attributes across the datasets.}
\label{fig:errors_datasets}
\end{figure}

\paragraph{Datasets} We evaluate our approach using three real-world datasets, which differ in size, number of errors, and number of considered constraints.  
Each dataset is accompanied by a gold standard that contains a sample of correct tuples. 
Table~\ref{tab:datasets} provides a summary of the dataset descriptions, including the the size of the gold standard and constraints considered in our experiments. 
\begin{itemize}
    \item \textbf{Allergen.} This dataset is composed of data about products and their allergens, which were crawled\footnote{Access date: 24 November, 2020.} from the German version of the Alnatura\footnote{\url{https://www.alnatura.de/de-de/}} and Open Food Facts\footnote{\url{https://world.openfoodfacts.org/}} websites. 
    The attributes of this dataset indicate the presence (`2'), traces (`1'), or absence (`0') of allergens in a product. 
    The constraints model the correspondence between allergens related to nuts in general and specific nuts.     
The gold standard was consolidated by the authors by looking at actual
pictures of the products' labels  provided by independent web sources.  
For the initial annotations, the inter-rater agreement computed with the Fleiss' kappa measure was $0.844$.
Discrepancies were resolved via mutual agreement. The repair of this dataset is particularly challenging due to the high number of attributes and error distribution across them. 

\item \textbf{EudraCT.} This dataset contains data about clinical trials conducted in Europe. This dataset was obtained from the European Union Drug Regulating Authorities Clinical Trials Database (EudraCT) register\footnote{\url{https://eudract.ema.europa.eu/}} and the gold standard was created from external registries\footnote{\url{https://clinicaltrials.gov/}}\footnote{\url{https://www.drks.de/drks_web/}}. 
The gold standard of this dataset does not contain the correct values for the attributes \textit{placebo} and \textit{active\_comparator}, hence, they do not appear in Figure~\ref{fig:errors_eudract}; yet, the edit rules $E_4$, $E_7$, $E_8$ and $E_9$ are kept as they contribute to the repairs of other attributes. 
The repair of this dataset is particularly challenging due to the number of constraints to consider. 

\item \textbf{Flight.}  This dataset describes flights annotated with the departing and arrival airports as well as their expected and actual time of departure and arrival. The dataset and gold standard used in this work are a modification of the Flight  Dataset\footnote{\url{http://lunadong.com/fusionDataSets.htm}}. We consolidated all the temporal information into the same timezone (UTC-0) and format. In this way, we can specify edit rules such as $E_1$ and $E_2$ (cf. Table~\ref{tab:datasets}). The repair of this dataset is challenging due to the high number of tuples. 
\end{itemize}   

\paragraph{Approaches} We compare our proposed solution Parker with recent state-of-the-art approaches for error correction using the following configurations. 

\begin{itemize}
    \item \textbf{Parker.} We evaluate different configurations of Parker (denoted $\cdot/\cdot/\cdot$) with variations in the key constraint $\fd$, the cost model $\cost$ and the repair selection strategy, respectively.
    For the \emph{key constraint}, we either consider a full key constraint (F) or no key constraint (N).
    We did not test a partial key scenario due to lack of golden standards for that configuration.
    The case with no key constraint allows to study an edit-rules-only case versus EPKs.
    For the \emph{cost model}, our baseline scenarios are constant cost functions (C) and cost functions that adopt an estimate of reliability based on observed errors (E) as introduced in Section~\ref{sec:cost-function-examples}.
    In the latter case, $\omega=4$ provided the best results.
    For the `Allergen' dataset, we additionally consider a preference cost function (P).
    This cost function models combines a reliability estimate ($\omega=4$) with a risk-aversion strategy, where we are reluctant to changes that remove an indication of an allergen (high cost) and are more inclined to accept repairs that add indications of allergens (low cost).
    For the \emph{repair selection}, we evaluate the frequency-based (F) and random (R) strategies.
    Frequency-based selection means here we sample repairs with a probability equal to their observed frequency in the clean data.
    In both cases, a frequent pattern tree was constructed from the clean data.
    As a final note, we also tested scenarios where we removed the edit rules and used only the key constraint. 
    Those scenarios are marked with a $*$ and in that case, we do not consider cost functions with a reliability estimate as there are no edit rules to detect tuple-level errors.
    
    \item \textbf{HoloClean~\cite{Rekatsinas2017}.}  We evaluate the performance of HoloClean\footnote{\url{https://github.com/HoloClean/holoclean} \texttt{v1.0.0}} using different error detector strategies: null detector (HoloClean-N), violation detector (HoloClean-V), and both detectors (HoloClean). The other parameters were used as provided by default.    
    
    \item \textbf{Raha+Baran~\cite{Mahdava2019,Mahdava2020}.} Raha and Baran\footnote{\url{https://github.com/BigDaMa/raha} \texttt{v.1.25}} are semi-supervised approaches that learn from correct labels. 
    Therefore, Raha and Baran were provided with the subset of each dataset for which the gold standard is available. 
    To detect errors, Raha relies on a labelling budget $\theta_{Label}$ to acquire correct labels for erroneous cells. We set up the labelling budget to  $\theta_{labels}=20$, as specified by the authors~\cite{Mahdava2019}.  
    The errors detected by Raha are then fed into Baran for correction. 
    We run Raha+Baran 10 times, and report on the average performance. 	
\end{itemize}

\paragraph{Evaluation Metrics} 
We report on precision, recall, and the $F_1$ metrics. 
A \textit{true positive} corresponds to a cell that is correctly repaired with respect to the gold standard.
Precision ($P$) is the number of correct repairs divided by the number of total repairs performed by the approach. 
Recall ($R$) is the number of correct repairs divided by the number of errors measured by comparing the input data with the gold standard. 
$F_1$ is the harmonic mean of precision and recall.  
We also measure the effectiveness of the approaches per attribute; we compute  precision, recall, and $F_1$ scores of the repairs in each attribute and average them by the number of attributes in the dataset to obtain the macro-precision ($P_M$), macro-recall ($R_M$) and macro-$F_1$ ($F_{1M}$).
The efficiency of the approaches is measured as the runtime in seconds ($s$). 

\paragraph{Implementation} Parker is implemented in Java 8 and is available as a part of the ledc-sigma\footnote{\url{https://gitlab.com/ledc/ledc-sigma}} package.
The implementation adopts also variables edit rules such as those appearing in Flight (Table~\ref{tab:datasets}).
HoloClean, Raha, and Baran are executed with Python 3.6. 
Parker and HoloClean use PostgreSQL 10.17 as backend. 
We execute the experiments on an Ubuntu 18.04.5 machine, with an i9-10940X CPU (3.30GHz, 14 cores with hyper-threading) and 256 GB RAM.
Execution times are measured with the Linux command \texttt{time}. \\

\subsection{Overall Effectiveness of the Approaches}
\label{sec:effectiveness}

\input{tables/results}
We compare the effectiveness of the repairs obtained with Parker and with the state-of-the-art approaches. 
Table~\ref{tab:effectiveness} presents a summary of the precision, recall, and $F_1$ scores of repairs for each dataset. 
Overall, we observe that none of the studied approaches completely overcomes the other baselines.  
This is due to the different repair strategies implemented by the approaches and the distribution of errors and coverage of edit rules in the datasets. 

The effectiveness of Parker depends on the cost model, the repair selection strategy, and the leverage of the constraints.  
In the Allergen dataset, the preferential cost model of Parker that implements a risk-aversion strategy clearly outperforms the other Parker configurations, since this type of model captures the intrinsic properties of this dataset.   
Yet, Parker can only handle repairs when the input sources disagree on the attribute values of an entity. 
For example, Parker could not compute correct repairs for the attribute \textit{cashew}, as both web sources indicated the absence of an allergen for the same product.  
This results in Parker achieving higher precision than recall. 
A similar behaviour is observed in Parker over the EudracCT dataset. 
In EudraCT, the error-based cost model (E) and frequency-based repair strategies (F) slightly outperform their corresponding counterparts with constant-based cost model (C) and random repairs (R), respectively.  
In addition, when edit rules are disabled in Parker (N), we observe a major drop in both precision and recall. 
This is due to the relatively high number of rows ($\sim10\%$) that are incorrect due to violations of edit rules; this aspect is further discussed in Section~\ref{sec:violation_results}.

In the Flight dataset, we tested only the configuration of Parker that selects repairs randomly (R).
Here, the error-based cost model (E) outperforms the other configurations of Parker. 
The reason for this is that E encodes the fact that sources with null values or violations of edit rules are considered less trustworthy, thus, increasing the cost of the repairs. 
To verify that this behaviour is not due to the effects of the edit rules, we also tested the additional configuration Parker F/E/R* (not reported in the table), which achieves close performance to F/E/R, i.e., 0.79 in F1-score, precision, and recall. 
This confirms that the gain in performance is mostly due to the error-based cost model (E) implemented by Parker, while the consideration of edit rules allow to correctly detect additional repairs.   

Comparing the best configurations of Parker across the different datasets, we observe that Parker performs better on EudraCT and Flight than on the Allergen dataset. 
The reason for this is the number of different sources describing the entities in each dataset. 
The Allergen dataset only includes two sources and, as explained earlier, Parker cannot detect repairs where the input sources agree on incorrect values. 
In contrast, the EudraCT and Flight datasets include several sources\footnote{Up to 44 sources in EudraCT and up to 51 in Flight.}. 
For these cases, the chances that all the sources carry the same errors are very small; this is leveraged by Parker to identify the repairs.   
Also, the configurations Parker F/C/F* and F/C/R* that ignore the edit rules do not outperform the other Parker configurations. 
This shows the advantage of exploiting the edit rules when computing dataset repairs. 
The impact of edit rules is further analyzed in Section~\ref{sec:violation_results}.  

In comparison to the other approaches, Parker exhibits high precision values, and competitive $F_1$ scores. 
The main reason for the lower recall values can be attributed to the cases where entities are described by a few sources with the same errors. 
In these cases, HoloClean and Raha+Baran can effectively learn repairs by taking into account the data distributions across the entire dataset, while Parker focuses on computing repairs at the entity level. 
Table~\ref{tab:effectiveness} also reports major differences between the performance of the state-of-the-art approaches. 
The results indicate that the learning techniques of HoloClean are able to learn repairs from larger datasets, as observed in the high precision values of HoloClean in the EudraCT and Flight datasets. 
Yet, HoloClean is not able to cover all possible errors in the data, which is reflected in the relatively low recall values in all datasets.    
In contrast, Raha+Baran can effectively learn repairs from smaller datasets but it still does not outperform Parker in terms of precision. 
From this, we can conclude that the learning techniques of Raha+Baran are recall-oriented allowing for identifying more types of repairs, while the cost model and the repair strategies of Parker are rather precision-oriented.       

\subsection{Violations of Edit Rules}
\label{sec:violation_results}
\input{tables/edit-rules-violations}

To further understand the effectiveness of the approaches, next we look into the violations to edit results defined for each dataset (cf. Table~\ref{tab:datasets}).     
Table~\ref{tab:rules_violations} shows the number of rows that violate the edit rules $E_i$ before and after computing repairs with the baseline approaches.  
This table distinguishes between the violations in the full dataset (left) and in the subset of the data for which the correct rows are available in the gold standard (right).
In comparison to the size of the datasets (cf. Table~\ref{tab:datasets}), we observe that the number of rows that violate edit rules is relatively low, i.e., $<1\%$ in Allergen, $11\%$ in EudraCT, and $22\%$ in Flight. 
This indicates that the high precision values achieved by Parker are not exclusively due to leveraging the edit rules, but also to its cost model and strategy to decide on the repairs.

The results in Table~\ref{tab:rules_violations} empirically confirm that Parker produces repairs that do not violate any of the edit rules in the datasets. 
Contrary, HoloClean and Raha+Baran are not able to ensure violation-free repairs and, in several cases, produce repairs that introduce more violations to edit rules.
In the case of HoloClean, despite that it takes into account edit rules for correcting the datasets, the computed repairs introduce new violations in the larger datasets EudraCT and Flight. 
For the EudraCT dataset, HoloClean introduces new violations for all rules except for $E_8$ and $E_9$.
Regarding the violations of $E_8$ and $E_9$, HoloClean did not produce repairs for this cells, i.e., the violations that occur in the original dataset were still present after the repairs.  
In the Flight dataset, the edit rules $E_1$ and $E_2$ cover all the the attributes of the dataset.
Therefore, we expect Parker and HoloClean to be able to detect the erroneous rows in the dataset that violate these rules.  
However, HoloClean introduces more violations, especially for the edit rule $E_1$, which involves the attributes $actual\_departure$ and $actual\_arrival$ with the highest number of errors according to Figure~\ref{fig:errors_flight}. 
This result indicates that HoloClean might learn incorrect repairs in the cases where the number of errors in some attributes is high. 
Raha+Baran only introduces new violations for the EudraCT dataset for the edit rules $E_1$, $E_4$, $E_7$, $E_8$, and $E_9$. 
In particular, the rules $E_4$, $E_7$, $E_8$, and $E_9$ involve the attributes $active\_comparator$ and $placebo$ which are not present in the gold standard as shown in Figure~\ref{fig:errors_eudract}. 
As expected, Raha+Baran is not able to produce repairs that do not violate rules for which the correct version is not available. 
In the other edit rules, Raha+Baran reduces the number of violations, yet, its effectiveness highly depends on the sampled tuples. 
For example, we observe that in $4$ out of the $10$ runs, Raha+Baran actually introduced new violations to edit rule EudraCT $E_5$, yet, it reduces the number of violations on average. 

Comparing the results of Tables~\ref{tab:effectiveness} and \ref{tab:rules_violations}, we confirm that the approaches that consider edit rules, i.e.,  Parker and HoloClean, achieve a higher precision in the rows where incorrect values are due to violations of edit rules. 

\subsection{Effectiveness of the Approaches per Attribute}
\label{sec:macro_effectiveness}
\begin{figure}[t!]
\begin{subfigure}[t]{0.99\textwidth}
\subcaption[c]{Allergen}
\label{fig:macro_allergen}
\includegraphics[width=\textwidth]{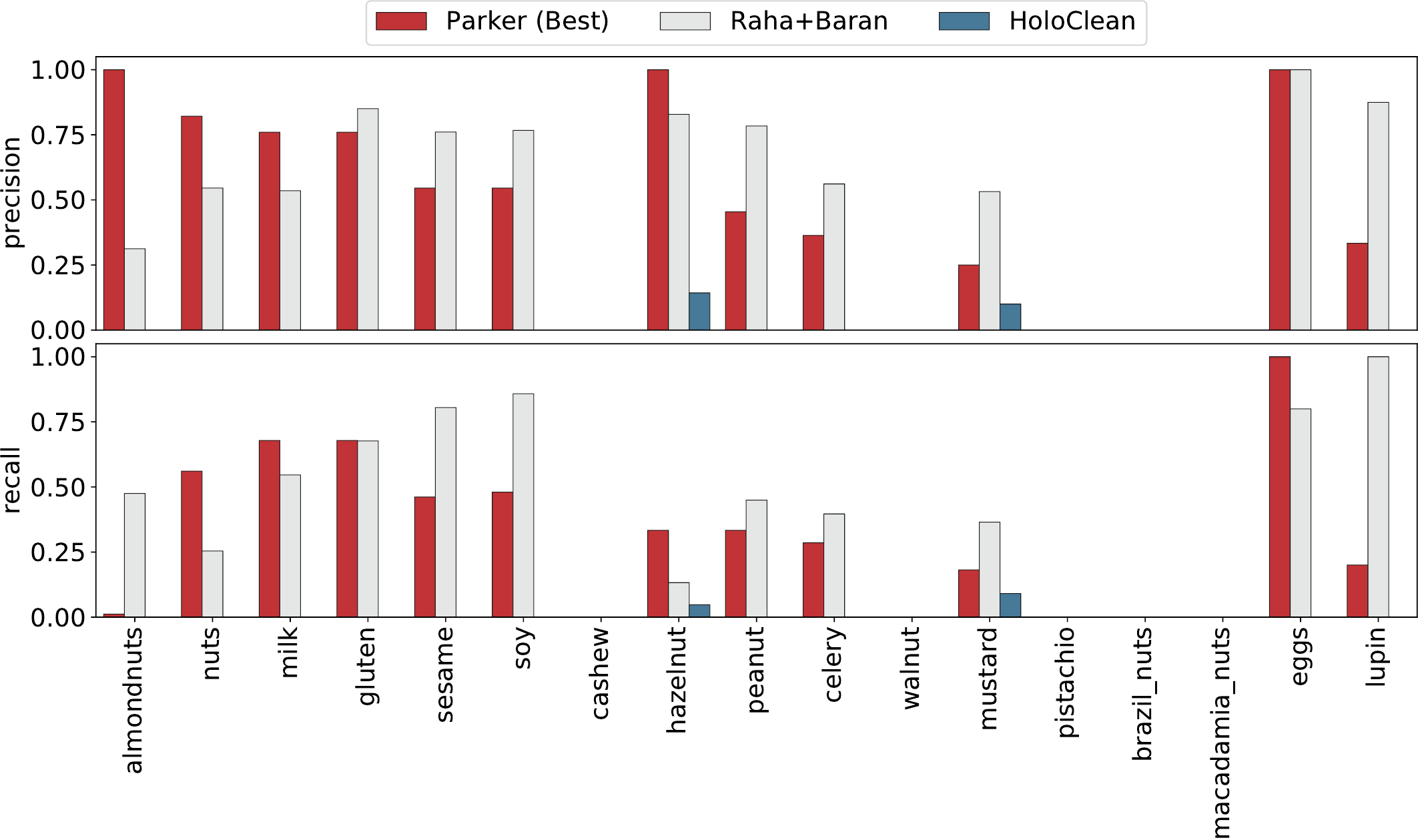}
\end{subfigure}

\vspace{1mm}

\begin{subfigure}[t]{0.66\textwidth}
\subcaption[c]{EudraCT}
\label{fig:macro_eudract}
\raisebox{18pt}{
\includegraphics[width=\textwidth]{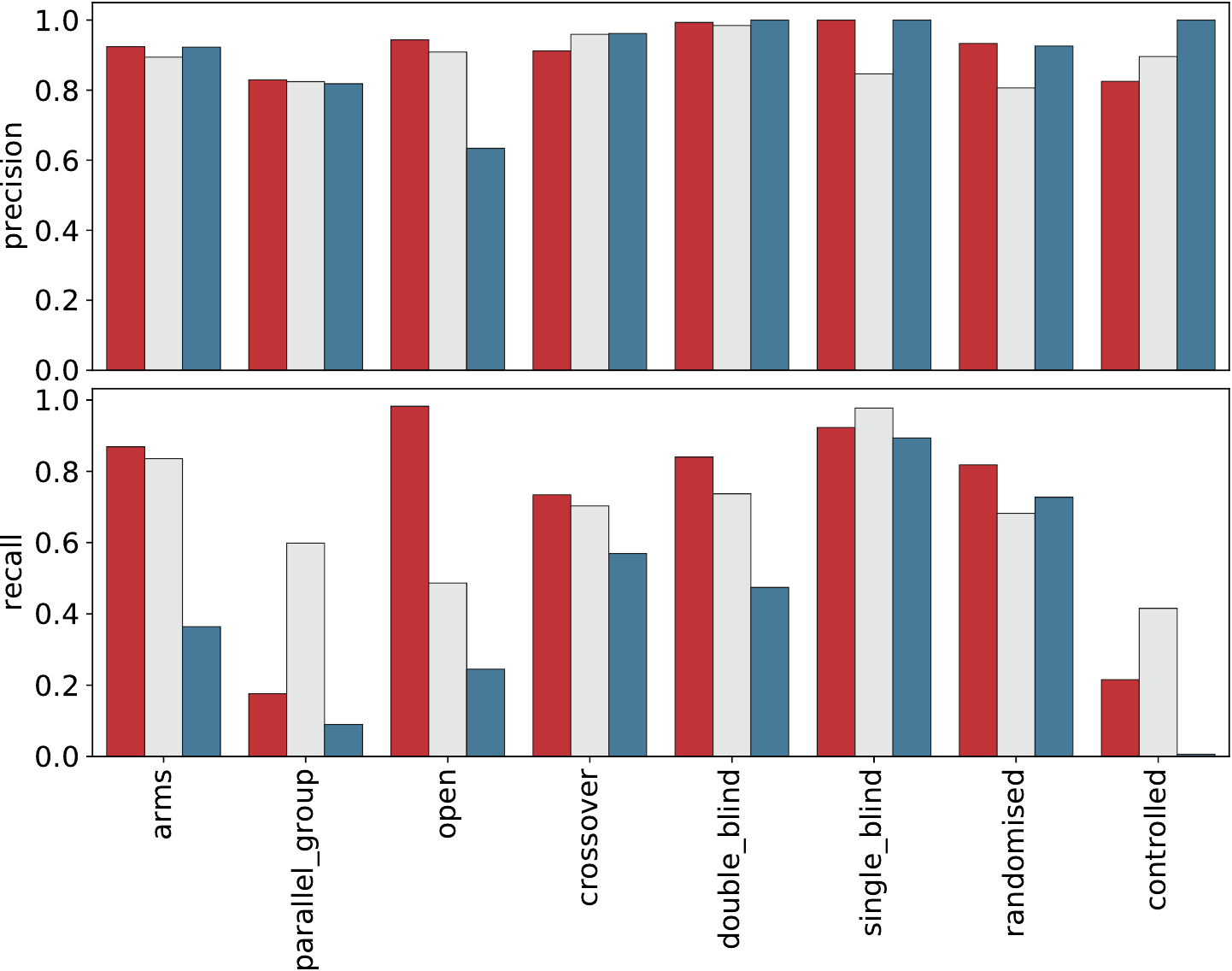}
}
\end{subfigure}
\hfill
\begin{subfigure}[t]{0.25\textwidth}
\subcaption[c]{Flight}
\label{fig:macro_flight}
\includegraphics[width=\textwidth]{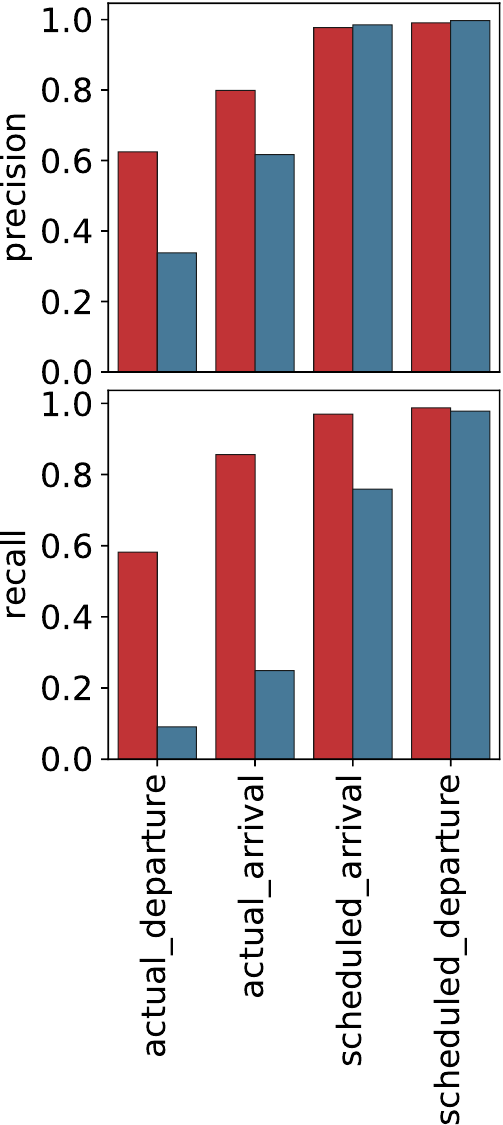}
\end{subfigure}
\caption{Precision and recall per attribute of repairs of the studied approaches. Attributes are ordered by the frequency of errors as of Figure~\ref{fig:errors_datasets}.}
\label{fig:performance_attribute}
\end{figure}

Figure~\ref{fig:errors_datasets} shows that the distribution of errors varies considerably among attributes within the datasets. 
Moreover, devising correct repairs for some attributes might be particularly challenging due to (i) the characteristics of some attributes, (ii) the skewness of error distributions, and (iii) the lack of edit rules covering the attributes. 
To provide further insights into these aspects of the repairs, in this section, we look into the effectiveness of the approaches in the different attributes. 
For each dataset, we first report on Table~\ref{tab:micro} on the macro scores for precision, recall, and $F_1$, where each metric is computed per attribute and then averaged by the number of attributes.  
Afterwards, we look into the precision and recall values for each attribute in Figure~\ref{fig:performance_attribute}.

\input{tables/macro-results}

\paragraph{Allergen Dataset} 
The macro scores in this dataset reveal that Raha+Baran outperforms the other approaches at the individual attributes. 
In comparison to the performance reported in Table~\ref{tab:effectiveness}, now Parker does not exhibit the highest precision. 
The reason for this is that Parker is able to identify correct repairs mainly for one of the attributes with the highest number of errors, i.e., $nuts$. 
Since all the edit rules $E_i$ cover this attribute (cf. Table~\ref{tab:datasets}), Parker effectively repairs  a large portion of the cells for the $nuts$ attribute. 
For most of the attributes, however, Parker does not have enough information to compute correct repairs as (i) there are no edit rules covering those attributes, and (ii) each entity is described by only two sources with potentially the same errors. 
This is similar for HoloClean, although it is not able to leverage effectively the edit rules for the attribute $nuts$. 
 In the case of Raha+Baran, the size of the dataset for which it is applied (206 rows) and the labelling budget $(\theta_{labels}=20)$ allows it to learn correct repairs for many more attributes that results in the best $P_M$ and $R_M$ performance. 
 Concretely, Raha+Baran outperforms Parker in $7$ out of $12$ attributes (cf.~Figure~\ref{fig:macro_allergen}). \\
 
\paragraph{EudraCT Dataset} 

In contrast to the overall results, Parker outperforms the state of the art in all the macro scores in this dataset. 
Table~\ref{tab:micro} shows that Raha+Baran performs slightly worse than Parker, indicating that the  learning techniques of Raha+Baran in combination with the labelling budget are not sufficient to capture all possible repairs in this dataset. 
In terms of $P_M$, most of the approaches achieve a relatively high precision in all attributes. 
Parker is only outperformed in the $crossover$ and $controlled$ attributes.  
Still, Parker is competitive with the other approaches in $crossover$, as this attribute presents errors that can be captured with edit rules ($E_5$ covers $17\%$ of the cases) or by cost model and repair strategies due to the high number of null values ($59\%$). 
%
The types of errors in the $controlled$ attribute are more challenging to detect and correct with Parker due to the low coverage with the edit rules and the lack of $null$ values. 
In these cases, using information about the data distributions as implemented by HoloClean allows for capturing more errors.       
In terms of $R_M$, Parker outperforms the state of the art in the majority of the attributes, expect for $single\_blind$, $controlled$, and $parallel\_group$. 
The attribute $single\_blind$ is covered by the edit rules $E_2$, $E_3$, and $E_4$ which are violated by $92\%$ of the rows. 
An inspection into the results for $single\_blind$ reveals that Parker correctly repaired all the rows that violate the edit rules, but it fails in correcting cells with other error types. 
For the attributes $controlled$ and $parallel\_group$, the edit rules $E_4$, $E_7$, and $E_5$ only cover $3\%$ and $5\%$ of the rows, respectively. 
This explains the low recall of Parker in these attributes.  
Interestingly, the recall of Parker is very different for $arms$ and $parallel\_group$, which are the attributes with the highest number of errors (cf. Figure~\ref{fig:errors_eudract}) and only a few rows ($< 3\%$) violate the edit rules covering these attributes.   
There is however a big difference in the amount of $null$ values: $91\%$ of the cells are $null$ for $arms$ vs. $16\%$ for $parallel\_group$.
At the same time, rules $E_8$ and $E_9$ allow for deductively repairing those studies that include either a placebo or an active comparator.
This shows how the presence of edit rules can lead to quite deterministic repairing in some cases.

\paragraph{Flight Dataset} 
Similar to the overall effectiveness of the approaches, Parker outperforms HoloClean in all the metrics (cf. Table~\ref{tab:micro}).  
Figure~\ref{fig:macro_flight} shows that the attributes $actual\_departure$ and $actual\_arrival$ are more challenging to repair for both approaches. 
The reason for this is two-fold: the edit rule $E_1$ that considers these attributes covers less rows than the other rule, and the number of $null$ values in these attributes is $40\%$ less than in the other attributes. 
In addition, the distribution of the errors shown in Figure~\ref{fig:errors_flight} suggests that it is challenging for HoloClean to learn correct repairs when a large portion of the rows contain errors;  
this is the case for $actual\_departure$ where $64\%$ of the cells are incorrect and $actual\_arrival$ where $51\%$ of the cells are incorrect.  
All these aspects make Parker, in general, more effective in both $P_M$ and $R_M$ in comparison to HoloClean.  \\

\subsection{Efficiency of the Approaches}
\label{sec:efficiency}
\input{tables/runtimes}

Lastly, we report the efficiency of the studied approaches in Table~\ref{tab:efficiency} in terms of runtime. 
Comparing all the approaches, Parker outperforms HoloClean and Raha+Baran by several orders of magnitude in all datasets.  
This confirms that Parker produces high-quality repairs which are competitive or more accurate than the state-of-the-art while requiring less computation time.   
In the following, we analyze the aspects that impact on the runtime of the  studied approaches.

For the learning-based approaches, i.e. HoloClean and Raha+Baran, we distinguish between the time spent during the error detection phase and the repair computation. 
For HoloClean and Parker, we exclude from this analysis the time required for loading the datasets into the database. 
Note that the results reported for Raha+Baran cannot be directly compared to the other approaches: HoloClean and Parker were executed over the entire datasets, while Raha+Baran was executed only on the subset of the datasets for which the rows exist in the gold standard. 

In the case of HoloClean, the error detection phase is directly impacted by the size of the datasets, but this effect is not observed for the computation of the repairs. 
For the latter stage, the performance is affected by the number of attributes and edit rules as HoloClean computes the distribution and correlations between attributes to learn the repairs. 

The performance of Raha+Baran follows a similar behavior to HoloClean. 
The error detection phase implemented by Raha is orders of magnitude faster than the repair of the cells with Baran. 
In particular, Raha+Baran is notably slower than HoloClean, as it is running only on a small percentage of the datasets ($3.6\% - 17.8\%$). 
The reason for this is the application of a large number of learning strategies and the computation of features to obtain the repairs.   
This affects the scalability of Raha+Baran to large datasets. 

Similar to HoloClean, the runtime of Parker is affected by the number of attributes and edit rules in the dataset. 
This is observed when comparing the runtimes for the configurations $\cdot$/$\cdot$/R for the datasets EudraCT and Flight: here,  Parker takes 3 times longer for EudraCT with a large number of attributes and edit rules than for  the Flight dataset with a large number of rows.  
In the case of Parker, Table~\ref{tab:efficiency} shows, as expected, that the frequency-based repair strategy is slower than the random strategy. 
Moreover, when the edit rules are not considered, the frequency-based strategy (configurations $\cdot$/$\cdot$/F*) requires more time than the random strategy ($\cdot$/$\cdot$/R*). 
This indicates that the edit rules allow for pruning the space of repairs, which improves both the accuracy of the approach and the efficiency. 
Overall, the runtimes of Parker indicate that it can scale up to even larger datasets.

\section{Conclusion and Future Work}
\label{sec:conclusion}
In this work, we presented a novel type of constraints called edit rules under a partial key (EPKs) to repair inconsistencies in data fusion. 
Based on EPKs, we then proposed an approach to compute minimal cost repairs which allow to efficiently and effectively explore the space of repairs. 
We demonstrated the theoretical properties regarding minimality of our approach.   
Our proposed techniques were implemented in our Parker engine. 
Our experimental study over three real-world datasets show that, on average, Parker computes more accurate repairs than the state-of-the-art approaches. 
In particular, Parker outperforms the state of the art in the cases where every entity is described by several sources and the correct repairs can be devised using the edit rules.     
In terms of efficiency, Parker showed to be orders of magnitude faster than the learning-based approaches. 
Our results indicate that Parker is able to scale up to very large datasets. 

Future work may focus on extending the computation of repairs by taking into consideration data distributions of the attributes. 
This will allow Parker to capture other repairs that are not covered by EPKs; yet, this would require further investigations on the tradeoff between accuracy and scalability of the computation of the repairs.  
Another line of work can investigate the application of our techniques to semantic data fusion, where additional ontological knowledge can be exploited to enhance the consistency constraints. 

\section{Appendix A: Proofs}
\begin{proof}[Proposition~\ref{prop:existence}]
If $\rules$ is satisfiable, there exists a tuple $\reptup$ that fails no rules.
A relation $\rel$ for which $\rel[\schema_1\cup\schema_2] = \reptup[\schema_1\cup\schema_2]$ is a non-empty relation that fails no EPKs.\qed
\end{proof}
\begin{proof}[Proposition~\ref{prop:subset-change}]
If a solution would not contain a $\subset$-minimal cover it would not cover failing rules and can thus never lead to a repair. \qed
\end{proof}
\begin{proof}[Theorem~\ref{theorem:boundedness}]
\boxed{1} By contradiction. On one hand, by construction and because any $\lcost{a}$ is positive definite, it follows that $\cost(r,r') = \cost(r[\cover], \reptup[\cover])$.
On the other hand, we have that $\cost(r,\reptup) = \cost(r[\cover], \reptup[\cover]) + \cost(r[\sol \setminus \cover], \reptup[\sol \setminus \cover])$.
Because $\reptup$ is a repair with solution $\sol$, the second term of the sum is greater than zero and it follows that $\cost(r,r')<\cost(r,\reptup)$.
So, if $r'$ is repair, then $\reptup$ is not $\cost$-minimal.

\boxed{2} If there exists a $\cover$-minimal repair, say $r^{**}$ with a cost smaller than or equal to $\cost\left(r,r'\right)$ then we have $\cost\left(r,r^{**}\right)\leq \cost\left(r,r'\right) < \cost\left(r,\reptup\right)$ from which it follows that $\reptup$ is not $\cost$-minimal.\qed
\end{proof}
\begin{proof}[Proposition~\ref{prop:necessity}]
Suppose the proposition would not hold, then there exists an $a_i\in\sol$ for which either $r[a_i]=\reptup[a_i]$ or $\forall \rul\in\suf{\rules}: \reptup[a_i] \in \rul_i$.
In the first case, then the solution is not $\sol$ but $\sol\setminus\{a_i\}$.
In the second case, consider the tuple $r'$ that satisfies $r'[\schema\setminus\{a_i\}] = \reptup[\schema\setminus\{a_i\}]$ and $r'[a_i]=r[a_i]$.
We now have $r'\models \rules$ and because cost functions are positive definite, we also have $\cost(r,r')<\cost(r,\reptup)$ which means $\reptup$ is not a $\cost$-minimal repair.
\qed
\end{proof}
\begin{proof}[Proposition~\ref{prop:indep2}]
For any $r\in\rel$, $r$ is modified into $\reptup$ with cost $\cost(r,\reptup)$.
Because $\rules'$ and $\rules''$ are independent sets, we have $\schema'\cup\schema'' = \emptyset$, so we have:
\begin{equation}
\cost(r,\reptup) = \cost(r[\kpart \cup\schema'],\reptup[\kpart \cup\schema']) + \cost(r[\kpart \cup\schema''],\reptup[\kpart \cup\schema'']).    
\end{equation}
Moreover, if $r'$ is a tuple with schema $\kpart \cup\schema'$ and $t$ is a tuple with schema $\schema$ then we have  $r'\models\rules' \wedge t[\schema'] = r' \Rightarrow t \models \rules'$.
A similar property holds for tuples $r''$ with schema $\kpart \cup\schema''$.
It then follows that if $r'$ is a tuple with schema $\kpart \cup\schema'$ and $r''$ is a tuple with schema $\kpart \cup\schema''$ we have:
\begin{equation}
r'\models\rules' \wedge r' \models \rules'' \Rightarrow r'\bowtie r'' \models \rules'\cup\rules'' = \rules.
\end{equation}
Any tuple $r'$ (with schema $\kpart \cup\schema'$) that satisfies $\rules'$ can thus be naturally joined with any tuple $r''$ (with schema $\kpart \cup\schema''$) that satisfies $\rules''$ and the result will satisfy $\rules$.
If in addition $r'$ minimizes $\cost(r[\kpart \cup\schema'],\reptup[\kpart \cup\schema'])$ and $r''$ minimizes $\cost(r[\kpart \cup\schema''],\reptup[\kpart \cup\schema''])$, then $r'\bowtie r''$ minimizes $\cost(r,\reptup)$.
\qed
\end{proof}
\begin{proof}[Theorem~\ref{theor:main}]
\boxed{1} If some tuple $r'\in M_\kvalue$ is consistent, then that tuple is a $\cost$-minimal repair for $\clas\subseteq \rel$ and also a trivial $\icost$-minimal repair for itself.

\boxed{2} If $\forall r'\in M_\kvalue: r'\not\models\rules$ then $\reptup\notin M_\kvalue$.
Now, for any $r''\notin M_\kvalue$ there always exists some tuple $t$ such that $\beta(\kvalue, t)<\beta(\kvalue, r'')$ and in addition, $r''$ and $t$ only differ for one attribute $a$.
We then have $\beta(\kvalue, r'') = \beta(\kvalue, t) + c(r''[a]) - c(t[a])$.
Now if $t\in M_\kvalue$, then we can write $\beta(\kvalue, r'') = \beta(\kvalue, t) + \icost(t,r'')$.
If $t\notin M_\kvalue$, then $t$ itself can be written in terms of some $t'$ where $t$ and $t'$ differ in only one attribute $b$ such that $\beta(\kvalue, r'') = \beta(\kvalue, t') + c(t[b]) - c(t'[b]) + c(r''[a]) - c(t[a])$.
Clearly, in each step $\beta(\kvalue, .)$ is strictly decreasing and lower bounded, so eventually, for every $r''\notin M_\kvalue$, we must find some $r'\in M_\kvalue$ such that $\beta(\kvalue, r'') = \beta(\kvalue, r') + \icost(r',r'')$.
It is now straightforward to see that any $\reptup\notin M_\kvalue$ can be written in the same way and minimizing $\beta(\kvalue, \reptup)$ is equivalent to minimizing $ \icost(r',\reptup)$ for any $r'\in M_\kvalue$.\qed
\end{proof}
\bibliographystyle{elsarticle-num-names}
\bibliography{references.bib}

\end{document}

%% file: tables/datasets.tex
\begin{table}[h!]
\caption{Dataset descriptions. The dataset size indicates the number of rows and attributes. Errors is the number of incorrect cells according to the Gold Standard (GS). The GS size includes the number (and percentage) of rows  covered from the dataset.} 
\centering\renewcommand\cellalign{lc}
\setcellgapes{0pt}\makegapedcells
\scriptsize
\begin{tabular}{l|l|r|r|l}
\toprule
\textbf{Dataset} & \textbf{Size} & \textbf{Errors} & \textbf{GS Size} & \textbf{Constraints}\\
\midrule
Allergen & $1160\times 22$ & $358$ & \makecell{$206$  \\$(17.8\%)$}  & 
\makecell{
\tiny{key = $code$}\\
\tiny{$E_1 = \neg(t_x.nuts < t_x.brazil\_nuts$)}  \\
\tiny{$E_2 = \neg(t_x.nuts < t_x.macadamia\_nuts)$} \\
\tiny{$E_3 = \neg(t_x.nuts < t_x.hazelnut)$} \\
\tiny{$E_4 = \neg(t_x.nuts < t_x.pistachio)$} \\
\tiny{$E_5 = \neg(t_x.nuts < t_x.walnut)$}
}\\
\midrule
EudraCT & $86670 \times 9$ & $2962$ & \makecell{$3133$ \\ $(3.6\%)$} & 
\makecell{
\tiny{key = $eudract\_number$}\\
\tiny{$E_1 = \neg(t_x.double\_blind = yes \wedge t_x.open = yes$)} \\
\tiny{$E_2 = \neg(t_x.single\_blind = yes \wedge t_x.open = yes$)} \\
\tiny{$E_3 = \neg(t_x.double\_blind = no \wedge t_x.single\_blind = no \wedge t_x.open = no$)} \\
\tiny{$E_4 = \neg(t_x.controlled = no \wedge t_x.placebo = yes$)}  \\ 
\tiny{$E_5 = \neg(t_x.crossover = yes \wedge t_x.parallel\_group = yes$)}  \\
\tiny{$E_6 = \neg(t_x.double\_blind = yes \wedge t_x.single\_blind = yes$)} \\
\tiny{$E_7 = \neg(t_x.active\_comparator = yes \wedge t_x.controlled = no$)} \\ 
\tiny{$E_8 = \neg(t_x.arms \leq 1 \wedge t_x.placebo = yes$)} \\ 
\tiny{$E_9 = \neg(t_x.arms \leq 1 \wedge t_x.active\_comparator = yes$)} \\ 
}\\
\midrule
Flight & $776067 \times 5$ & $134778$ & \makecell{$70951$ \\ $(9\%)$} & 
\makecell{
\tiny{key = $\{date\_collected,flight\_number\}$}\\
\tiny{$E_1 = \neg(t_x.actual\_departure \ge t_x.actual\_arrival$)}  \\
\tiny{$E_2 = \neg(t_x.scheduled\_departure \ge t_x.scheduled\_arrival$)} 
}\\
\bottomrule
\end{tabular}
\label{tab:datasets}
\end{table}

%% file: tables/results.tex
\begin{table}[t!]
\caption{Effectiveness of the repairs of the evaluated approaches. \textsc{n/a} indicates that the approach configuration is not applicable to that dataset. -- indicates that the approach could not be executed in that dataset. Best results per dataset are highlighted in bold.}
\label{tab:effectiveness}
\centering
\footnotesize
\begin{tabular}{l|rrr|rrr|rrr}
\toprule
& \multicolumn{3}{c|}{\textbf{Allergen}}& \multicolumn{3}{c|}{\textbf{EudraCT}} & \multicolumn{3}{c}{\textbf{Flight}}\\
\textbf{Approach}& \multicolumn{1}{c}{$F_1$} & \multicolumn{1}{c}{$P$} & \multicolumn{1}{c}{$R$} & \multicolumn{1}{|c}{$F_1$} & \multicolumn{1}{c}{$P$} & \multicolumn{1}{c}{$R$} & \multicolumn{1}{|c}{$F_1$} & \multicolumn{1}{c}{$P$} & \multicolumn{1}{c}{$R$} \\
\midrule
HoloClean& 0.01& 0.06& 0.01& 0.49& 0.91& 0.33& 0.56& 0.81& 0.43\\
HoloClean-N & \textsc{n/a} & \textsc{n/a} & \textsc{n/a} & 0.47& 0.87& 0.32& 0.57& 0.81& 0.43\\
HoloClean-V & 0.01& 0.06& 0.01& 0.49& 0.91& 0.33& 0.56& 0.81& 0.43\\

\midrule

Raha+Baran	&\textbf{0.46}	&0.51	& \textbf{0.42}	& \textbf{0.77} 	&0.87	&\textbf{0.70}&	--	& -- 	& -- \\ 
\midrule
Parker F/E/F  & 0.13& 0.20& 0.09& 0.76& \textbf{0.93}& 0.65&  \textsc{n/a} & \textsc{n/a} & \textsc{n/a}\\
Parker F/C/F  & 0.15& 0.24& 0.11& 0.74& 0.90& 0.62& \textsc{n/a} & \textsc{n/a} & \textsc{n/a} \\
Parker F/E/R  & 0.29& 0.45& 0.21& 0.65& 0.80& 0.55& \textbf{0.81}& \textbf{0.82} & \textbf{0.81} \\
Parker F/C/R  & 0.28& 0.44& 0.20& 0.64& 0.79& 0.54& 0.68& 0.69& 0.67\\
Parker N/C/F  & 0.01& 0.50& 0.01& 0.50& 0.63& 0.41& \textsc{n/a} & \textsc{n/a} & \textsc{n/a} \\
Parker N/C/R  & 0.01& 0.50& 0.01& 0.48& 0.60& 0.40& 0.46 & 0.44 & 0.49 \\
Parker F/C/F*  & 0.14& 0.22& 0.10& 0.69& 0.88& 0.57& \textsc{n/a} & \textsc{n/a} & \textsc{n/a} \\
Parker F/C/R*  & 0.26& 0.40& 0.19& 0.57& 0.73& 0.47& 0.68& 0.70& 0.67\\
Parker F/P/F  & 0.42& \textbf{0.65}& 0.31& \textsc{n/a} & \textsc{n/a} & \textsc{n/a} & \textsc{n/a} & \textsc{n/a} & \textsc{n/a}  \\
\bottomrule
\end{tabular}
\end{table}

%% file: tables/edit-rules-violations.tex
\begin{table}[t!]
\caption{Violations of edit rules. For each edit rule $E_i$, number of rows (with non-null values in the involved attributes) in the dataset and after the repairs that violate $E_i$. The right part of the table focuses on the subset of the dataset for which exists an entry in the Gold Standard (GS). Parker (All) indicates all configurations. For Raha+Baran, the table reports the average number of violations obtained in the $10$ runs.}
\label{tab:rules_violations}
\scriptsize
\begin{tabular}{r|rrr|rrrr}
\toprule
\textbf{Edit Rule}    & \textbf{Dataset} & \textbf{HoloClean} & \textbf{Parker} & \textbf{Subset} & \textbf{HoloClean} & \textbf{Raha+} & \textbf{Parker} \\
\textbf{}    & \textbf{} & \textbf{} & \textbf{(All)} & \textbf{in GS} & \textbf{} & \textbf{Baran} & \textbf{(All)} \\
\midrule
{Allergen $E_1$} & 0    & 0    & 0    & 0    & 0    & 0     & 0    \\
{Allergen $E_2$} & 0    & 0    & 0    & 0    & 0    & 0     & 0    \\
{Allergen $E_3$} & 10    & 1    & 0    & 3    & 0    & 2.4     & 0    \\
{Allergen $E_4$} & 0    & 0    & 0    & 0    & 0    & 0     & 0    \\
{Allergen $E_5$} & 0    & 0    & 0    & 0    & 0    & 0     & 0    \\
{Total Violations} & 10    & 1    & 0    & 3    & 0    & 2.4     & 0    \\
\midrule
{EudraCT $E_1$} & 881   & {\color[HTML]{c13336}895}    & 0    & 12   & 12    & {\color[HTML]{c13336}12.6}    & 0    \\
{EudraCT $E_2$} & 164   & {\color[HTML]{c13336}167}    & 0    & 0    & 0    & 0     & 0    \\
{EudraCT $E_3$} & 7034   & {\color[HTML]{c13336}7953}    & 0    & 151   & {\color[HTML]{c13336}192}    & {\color[HTML]{c13336}180.9}    & 0    \\
{EudraCT $E_4$} & 57    & {\color[HTML]{c13336}65}    & 0    & 0    & 0    & {\color[HTML]{c13336}4.5}     & 0    \\
{EudraCT $E_5$} & 890   & {\color[HTML]{c13336}896}    & 0    & 47   & 47    & 28.8    & 0    \\
{EudraCT $E_6$} & 211   & {\color[HTML]{c13336}217}    & 0    & 5    & 5    & 0     & 0    \\
{EudraCT $E_7$} & 165   & {\color[HTML]{c13336}190}    & 0    & 5    & {\color[HTML]{c13336}7}    & {\color[HTML]{c13336}16.7}    & 0    \\
{EudraCT $E_8$} & 230   & 230    & 0    & 2    & 2    & {\color[HTML]{c13336}2.5}     & 0    \\
{EudraCT $E_9$} & 146   & 146    & 0    & 7    & 7    & {\color[HTML]{c13336}8.9}     & 0    \\
{Total Violations} & 9778   & {\color[HTML]{c13336}10759}    & 0    & 229    & {\color[HTML]{c13336}272}    & {\color[HTML]{c13336}254.9}     & 0    \\
\midrule
{Flight $E_1$} & 9327    & {\color[HTML]{c13336}21375}    & 0    & 799    & {\color[HTML]{c13336}1861}    & --     & 0    \\
{Flight $E_2$} & 10055   & {\color[HTML]{c13336}10640}    & 0    & 1273   & {\color[HTML]{c13336}1367}    & --     & 0 \\ 
{Total Violations} & 19382   & {\color[HTML]{c13336}32015}    & 0    & 2072   & {\color[HTML]{c13336}3228}    & --     & 0 \\   
\bottomrule
\end{tabular}
\end{table}

%% file: tables/macro-results.tex
\begin{table}[t!]
\caption{Macro scores of the repairs of the evaluated approaches. -- indicates that the approach could not be executed in that dataset. Parker (Best) indicates the best configuration per dataset as of Table~\ref{tab:effectiveness}, i.e., F/P/F for Allergen and F/CE/F for EudraCT and Flight.  Best results per dataset are highlighted.}
\label{tab:micro}
\centering
\footnotesize
\begin{tabular}{l|rrr|rrr|rrr}
\toprule
            & \multicolumn{3}{c|}{\textbf{Allergen}}                                                            & \multicolumn{3}{c|}{\textbf{EudraCT}}                                                             & \multicolumn{3}{c}{\textbf{Flight}}                                                            \\
\textbf{Approach}            & \multicolumn{1}{c}{$F_{1M}$} & \multicolumn{1}{c}{$P_M$} & \multicolumn{1}{c}{$R_M$} & \multicolumn{1}{|c}{$F_{1M}$} & \multicolumn{1}{c}{$P_M$} & \multicolumn{1}{c}{$R_M$} & \multicolumn{1}{|c}{$F_{1M}$} & \multicolumn{1}{c}{$P_M$} & \multicolumn{1}{c}{$R_M$} \\
\midrule
HoloClean                      & 0.01                           & 0.05                           & 0.01                           & 0.58                           & 0.91                           & 0.42                           & 0.61                           & 0.73                           & 0.52                           \\
Raha+Baran & \textbf{0.51}                           & \textbf{0.69}                           & \textbf{0.40}                           & 0.77                           & 0.89                           & 0.68                           & --                             & --                             & --                             \\
Parker (Best)                       & 0.42                           & 0.65                           & 0.31                           & \textbf{0.79}                           & \textbf{0.92}                           & \textbf{0.69}                           & \textbf{0.85}                           & \textbf{0.85}                           & \textbf{0.85} \\
\bottomrule                         
\end{tabular}

\end{table}

%% file: tables/runtimes.tex
\begin{table}[t!]
\caption{Runtime of the approaches (in seconds). For the learning-based approaches, the runtimes for error detection and repair computation are reported. For Parker, average runtimes of the configurations that implement a frequency-based (F) or a random-based (R) repair strategy. * indicates that edit rules are ignored.}
\label{tab:efficiency}
\setlength{\tabcolsep}{10pt}
\centering
\footnotesize
\begin{tabular}{lrrr}
\toprule
\textbf{Approach}        & \textbf{Allergen} & \textbf{EudraCT} & \textbf{Flight}   \\
\midrule
HoloClean Error Detection & 0.87     & 22.47   & 462.03  \\
HoloClean Repairs         & 115.61   & 6099.57 & 6841.80 \\
HoloClean Total           & 116.48   & 6122.04 & 7303.83 \\
\midrule
Raha Error Detection            & 8.31     &  33.80       & --      \\
Baran Repairs            & 326.95   & 5493.22        & --      \\
Raha+Baran Total           & 335.25   & 5527.02        & --      \\
\midrule
Parker $\cdot$/$\cdot$/F           & 0.005     & 8.95    & \textsc{n/a}      \\
Parker $\cdot$/$\cdot$/R           & 0.005    & 6.00    & 2.02    \\
Parker $\cdot$/$\cdot$/F*          & 0.003     & 31.31   & \textsc{n/a}      \\
Parker $\cdot$/$\cdot$/R*          & 0.003     & 1.93    & 1.87   \\
\bottomrule 
\end{tabular}
\end{table}